\pdfoutput=1
\RequirePackage{ifpdf}
\ifpdf 
\documentclass[pdftex]{sigma}
\else
\documentclass{sigma}
\fi

\def \arctanh{\mathop{\rm arctanh}\nolimits}
\def \arccoth{\mathop{\rm arccoth}\nolimits}
\def \cotan{\mathop{\rm cotan}\nolimits}
\def \sech{\mathop{\rm sech}\nolimits}

\numberwithin{equation}{section}

\begin{document}

\allowdisplaybreaks

\renewcommand{\thefootnote}{$\star$}

\renewcommand{\PaperNumber}{017}

\FirstPageHeading

\ShortArticleName{Solitary Waves in Massive Nonlinear $\mathbb{S}^N$-Sigma Models}

\ArticleName{Solitary Waves in Massive\\ Nonlinear $\boldsymbol{\mathbb{S}^N}$-Sigma Models\footnote{This
paper is a contribution to the Proceedings of the Eighth
International Conference ``Symmetry in Nonlinear Mathematical
Physics'' (June 21--27, 2009, Kyiv, Ukraine). The full collection
is available at
\href{http://www.emis.de/journals/SIGMA/symmetry2009.html}{http://www.emis.de/journals/SIGMA/symmetry2009.html}}}

\Author{Alberto ALONSO IZQUIERDO~$^\dag$, Miguel \'Angel GONZ\'ALEZ LE\'ON~$^\dag$\\
 and Marina DE LA TORRE MAYADO~$^\ddag$}

\AuthorNameForHeading{A. Alonso Izquierdo, \'M.A. Gonz\'alez Le\'on and M. de la Torre Mayado}

\Address{$^\dag$~Departamento de Matem\'atica Aplicada,  Universidad de Salamanca, Spain}
\EmailD{\href{mailto:alonsoiz@usal.es}{alonsoiz@usal.es}, \href{mailto:magleon@usal.es}{magleon@usal.es}}
\URLaddressD{\url{http://campus.usal.es/~mpg/}}

\Address{$^\ddag$~Departamento de F\'{\i}sica Fundamental, Universidad de Salamanca, Spain}
\EmailD{\href{mailto:marina@usal.es}{marina@usal.es}}

\ArticleDates{Received December 07, 2009;  Published online February 09, 2010}

\Abstract{The solitary waves of massive $(1+1)$-dimensional
nonlinear ${\mathbb S}^N$-sigma models are unveiled. It is shown
that the solitary waves in these systems are in one-to-one correspondence with
the separatrix trajectories  in the repulsive $N$-dimensional Neumann mechanical problem.
There are topological (heteroclinic trajectories) and non-topological (homoclinic trajectories)
kinks. The stability of some embedded sine-Gordon kinks is discussed by means of the
direct estimation of the spectra of the second-order f\/luctuation
operators around them, whereas the instability of other topological and
non-topological kinks  is established applying the Morse index
theorem.}


\Classification{35Q51;  81T99}

\renewcommand{\thefootnote}{\arabic{footnote}}
\setcounter{footnote}{0}

\section{Introduction}

The existence and the study of solitary waves in $(1+1)$-dimensional
relativistic f\/ield theories is
of the greatest interest in several scientif\/ic domains. We specif\/ically
mention high energy physics, applied mathematics, condensed matter physics, cosmology,
f\/luid dynamics, semiconductor physics, etcetera. Solitary waves
have been discovered recently in massive nonlinear sigma models with quite striking properties,
see e.g.~\cite{PRL, PRD,JPA} and references quoted
therein. The nonlinear ${\mathbb S}^N$-sigma models are important in describing the low energy
pion dynamics in nuclear physics~\cite{Gell-Mann/Levy} and/or featuring the long wavelength behavior of
${\mathbb O}(N+1)$-spin chains in ferromagnetic materials~\cite{Brezin/Zinn-Justin}.

In this work, we present a mathematically detailed analysis of the
solitary wave manifold arising in  a massive ${\mathbb S}^N$-sigma model
that extends and generalize previous work performed in~\cite{PRL} and
\cite{PRD} on the $N=2$ case. In these papers the masses of the $N=2$ meson branches
are chosen to be dif\/ferent by breaking the ${\mathbb O}(N)$ degeneracy of the ground
state to a ${\mathbb Z}_2$ sub-group. Remarkably, we recognized in~\cite{PRL}
and \cite{PRD} that the static f\/ield equations of this non-isotropic massive
nonlinear ${\mathbb S}^2$-sigma model are the massive version of the static
Landau--Lifshitz equations governing the high spin and long
wavelength limit of 1D ferromagnetic materials. This perspective
allows to interpret the topological kinks of the model as Bloch and
Ising walls that form interfaces between ferromagnetic domains, see~\cite{BW, BW1,BWZ}.

We shall proceed into a two-step mathematical analysis: First, we shall identify the collection
of $N$ sine-Gordon models embedded in our nonlinear ${\mathbb S}^N$-sigma models. The
sine-Gordon kinks (and the sine-Gordon multi-solitons) are automatically solitary waves (and multi-solitons)
in the nonlinear sigma model. Second, we shall identify the
static f\/ield-equations as the Newton equations of the repulsive Neumann problem in the
${\mathbb S}^N$-sphere. The mechanical analogy in the
search for solitary waves in $(1+1)$-dimensional relativistic f\/ield theories, \cite{Rajaraman}, is
the re-reading of the static f\/ield equations as the motion equations of a
mechanical system where the f\/ield theoretical spatial coordinate is
re-interpreted as the ``mechanical time'' and the potential energy density is minus the
``mechanical potential''. This strategy is particularly ef\/f\/icient in f\/ield theories with a single
scalar f\/ield because in such a case the mechanical analogue problem has only one
degree of freedom and one can always integrate the motion equations.

The mechanical analogue system of f\/ield theoretical models with $N$ scalar f\/ields has $N$
degrees of freedom and the system of $N$ motion ODE's is rarely integrable. One may analytically f\/ind
the solitary waves in f\/ield theories such that their associated mechanical system is completely integrable.
Many solitary waves have been found in two scalar f\/ield models related to the Garnier integrable
system of two degrees of freedom and other Liouville models, see \cite{AMAJ}. The extremely rich structure
of such varieties of kinks has been exhaustively described in \cite{AJ}. The procedure can be also
extended to systems with $N=3$ (and higher) degrees of freedom that are Hamilton--Jacobi separable using Jacobi
elliptic coordinates, see \cite{Nonlinearity1}. The mechanical analogue
system of the non-isotropic massive nonlinear ${\mathbb S}^N$-system is the Neumann system, a particle
constrained to move in a ${\mathbb S}^N$-sphere subjected to elastic repulsive non-isotropic forces from
the origin. It happens that the Neumann system \cite{Neumann, Moser, Dubrovin} is not only Arnold--Liouville
completely integrable~-- there are $N$ constants of motion in involution~-- but Hamilton--Jacobi separable
using sphero-conical coordinates. Because the elastic forces are repulsive the North and South poles
are unstable equilibrium points where heteroclinic and homoclinic trajectories start and end
providing  the solitary waves of the nonlinear sigma model.

The analysis of the stability of solitary waves against small f\/luctuations is
tantamount to the study of the linearized wave equations around a particular
traveling wave. We shall distinguish between stable and unstable sine-Gordon kinks
by means of the determination of the spectra of small f\/luctuations around these embedded
kinks. The instability of any other (not sine-Gordon) solitary wave in the
nonlinear ${\mathbb S}^3$-sigma model will be shown by computing the
Jacobi f\/ields between conjugate points along the trajectory and using the Morse index theorem
within the framework explained in \cite{Nonlinearity}.

The structure of the article is as follows: In Section~\ref{section2} we present
the fundamental aspects of the model. Section~\ref{section3} is devoted to the
study of the set of sine-Gordon models that are embedded in the
nonlinear ${\mathbb S}^N$-sigma model. In Section~\ref{section4}, using sphero-conical
coordinates, we show the
separability of the Hamilton--Jacobi equation that reduces to $N$ uncoupled ODE.
The Hamilton characteristic function is subsequently found as the sum of $N$
functions, each one depending only on one coordinate, and the separation constants are chosen
in such a way that the mechanical energy is zero. Then, the motion equations are solved
by quadratures. The properties and structure of the explicit analytic solutions are discussed
respectively in Sections~\ref{section5} and~\ref{section6} for the $N=2$ and $N=3$ models.
Finally, in Section~\ref{section7} the stability analysis of the solitary wave
solutions is performed.

\section[The massive nonlinear ${\mathbb S}^N$-sigma model]{The massive nonlinear $\boldsymbol{{\mathbb S}^N}$-sigma model}\label{section2}

 Let us consider $N+1$ real scalar f\/ields arranged in a
vector-f\/ield:
\[
\vec{\Phi}\big(y^0,y^1\big)\equiv \big(\phi_1\big(y^0,y^1\big),\dots
,\phi_{N+1}\big(y^0,y^1\big)\big),
\]
where $\big(y^0,y^1\big)$ stand for the standard coordinates in the
Minkowski space ${\mathbb R}^{1,1}$, and the f\/ields are constrained
to be in the ${\mathbb S}^N$-sphere, i.e.
\begin{gather}
\vec{\Phi}  \in   {\rm Maps}   \left( {\mathbb R}^{1,1},{\mathbb
S}^{N}\right)  ,\qquad \vec{\Phi}\cdot \vec{\Phi}  =  \phi_1^2+
\dots  +  \phi_{N+1}^2  =  R^2  .\label{constraint}
\end{gather}
The dynamics of the model is governed by the action functional:
\begin{gather*}
S[\vec{\Phi}]  =  \int dy^0  dy^1   \,   {\cal L}\big(
\partial_\mu \vec{\Phi},\vec{\Phi}\big)  =  \int dy^0dy^1
\left\{ \frac{1}{2} \partial_\mu\vec{\Phi}\cdot \partial^\mu
\vec{\Phi}  -  V(\vec{\Phi})\right\},\label{action}
\end{gather*}
where $V(\vec{\Phi})$ is the quadratic polynomial function in the
f\/ields:
\begin{gather}
V(\vec{\Phi})=\tfrac{1}{2}  \left( \alpha_1^2
\phi_1^2+\alpha_2^2  \phi_2^2+  \dots  +\alpha_{N+1}^2
\phi_{N+1}^2-\alpha_{N+1}^2 R^2   \right)   .\label{potden}
\end{gather}
We follow standard Minkowski space conventions: $ y^\mu
y_\mu=g^{\mu\nu} y_\mu y_\nu$, $g^{\mu\nu}={\rm diag} (1,-1)$,
$\partial_\mu\partial^\mu
=g^{\mu\nu}\partial_\mu\partial_\nu=\partial_0^2-\partial_1^2$.
Working in the natural system of units, $\hbar=c=1$, the dimensions
of the f\/ields and parameters are: $[\phi_a]=[R]=1$, $[\alpha_a]=M$,
$a=1, \dots , N+1$.

The f\/ield equations of the system are:
\begin{gather}
\partial_\mu\left( \frac{\partial {\cal L}}{\partial
 (\partial_\mu\phi_a)}\right)-\frac{\partial{\cal L}}{\partial
 \phi_a}=0  \  \Rightarrow  \    \partial_0^2   \phi_a-\partial_1^2
 \phi_a=-\alpha_a^2   \phi_a+\lambda  \phi_a  ,\qquad
 a=1,\dots,N+1    ,\label{eqq0}
\end{gather}
$\lambda$ being the Lagrange multiplier associated to the constraint
(\ref{constraint}).  $\lambda$ can be expressed in terms of the
f\/ields and their f\/irst derivatives:
\begin{gather*}
\lambda=\frac{1}{R^2}\left( \sum_{a=1}^{N+1} \alpha_a^2
\phi_a^2-\partial_0\vec{\Phi}\cdot
\partial_0\vec{\Phi}+\partial_1\vec{\Phi}\cdot
\partial_1\vec{\Phi}\right)\label{lambda}
\end{gather*}
by successive dif\/ferentiations of the constraint equation
(\ref{constraint}). There exists $N+1$ static and homogeneous
solutions of (\ref{eqq0}) on ${\mathbb S}^{N}$:
\begin{gather}
\phi_a=\pm R  ,\qquad  \phi_b=0, \quad \forall \, b\neq a, \qquad
\lambda=\alpha_a^2  ,\qquad a=1,\dots,N+1
.\label{static-homogeneous}
\end{gather}
We shall address the maximally anisotropic case: $\alpha_a\neq
\alpha_b$, $\forall \, a\neq b$, and, without loss of generality, we
label the parameters in decreasing order: $ \alpha_1^2>\alpha_2^2>
\cdots   >\alpha_{N+1}^2\geq 0$, in such a~way that $V(\vec{\Phi})$,
on the ${\mathbb S}^N$-sphere (\ref{constraint}), is semi-def\/inite
positive.

According to Rajaraman \cite{Rajaraman} solitary waves (kinks) are
non-singular solutions of the f\/ield equations (\ref{eqq0}) of f\/inite
energy such that their energy density has a space-time dependence of
the form: $\varepsilon(y^0,y^1)=\varepsilon(y^1-vy^0)$, where $v$ is
some velocity vector. The energy functional is:
\begin{gather*}
E[\vec{\Phi}]=\displaystyle  \int dy^1 \left( \frac{1}{2}
\partial_0\vec{\Phi}\cdot \partial_0 \vec{\Phi}+
\frac{1}{2}   \partial_1\vec{\Phi}\cdot \partial_1
\vec{\Phi}+V\big(\vec{\Phi}\big)\right)   =  \int dy^1
\varepsilon\big(y^0,y^1\big)
\end{gather*}
and the integrand $\varepsilon\big(y^0,y^1\big)$ is the energy density. Lorentz invariance
of the model implies that it suf\/f\/ices to know the $y^0$-independent
solutions $\vec{\Phi}\big(y^1\big)$ in order to obtain
the traveling waves of the model:
$\vec{\Phi}\big(y^0,y^1\big)=\vec{\Phi}\big(y^1-vy^0\big)$. For static
conf\/igurations the energy functional reduces to:
\begin{gather}
E[\vec{\Phi}]=\int dy^1 \left( \frac{1}{2}   \frac{d
\vec{\Phi}}{dy^1}\cdot \frac{d\vec{\Phi}}{dy^1}  +
V\big(\vec{\Phi}\big)\right)  =  \int dy^1     \varepsilon\big(y^1\big)
.\label{energy}
\end{gather}
and the PDE system (\ref{eqq0}) becomes the following system of $N+1$
ordinary dif\/ferential equations:
\begin{gather}
\frac{d^2 \phi_a}{d(y^1)^2}  =  \alpha_a^2   \phi_a -\lambda
\phi_a. \label{eqqq}
\end{gather}

The f\/inite energy requirement selects between the static and
homogeneous solutions (\ref{static-homogeneous}) only two, the set
${\cal M}$ of zeroes (and absolute minima) of $V(\vec{\Phi})$ on
${\mathbb S}^N$:
\begin{gather*}
{\cal M}  =  \left\{  v^+=(0,\dots,0, R)  ,\  v^-=(0,\dots,0,-
R)\right\} ,\label{zeroes}
\end{gather*}
i.e.\ the ``North'' and the ``South'' poles of the ${\mathbb
S}^{N}$-sphere. Moreover, this requirement forces in~(\ref{energy})
the asymptotic conditions:
\begin{gather}
\lim_{y^1\to \pm \infty}   \frac{d \vec{\Phi}}{dy^1}  =  0
,\qquad \lim_{y^1\to \pm \infty}   \vec{\Phi}  \in {\cal M}    .\label{asy}
\end{gather}
Consequently, the conf\/iguration space, the space of f\/inite energy conf\/igurations:
\[
{\cal C}=\left\{{\rm Maps}({\mathbb R},{\mathbb S}^{n})/ E<+\infty \right\},
\]
is the union of four disconnected sectors: $ {\cal C}={\cal C}_{\rm
NN}\bigcup {\cal C}_{\rm SS} \bigcup {\cal C}_{\rm NS} \bigcup {\cal
C}_{\rm SN}$, where the dif\/ferent sectors are labeled by the element
of ${\cal M}$ reached by each conf\/iguration at $y^1\to -\infty$ and
$y^1\to \infty$. ${\cal C}_{\rm NS}$~and~${\cal C}_{\rm SN}$
solitary waves will be termed as topological kinks, whereas
non-topological kinks belong to ${\cal C}_{\rm NN}$ or ${\cal
C}_{\rm SS}$.

\section{Embedded sine-Gordon models}\label{section3}

Before of embarking ourselves in the general solution of the ODE system (\ref{eqqq}),
we explore Rajaraman's trial-orbit method \cite{Rajaraman}. We
search for solitary wave solutions living on special meridians of
the ${\mathbb S}^N$-sphere such that the associated trajectories connect the North and South poles,
to comply with the asymptotic conditions of f\/inite energy (\ref{asy}).
In \cite{PRL} and \cite{PRD} this strategy was tested in the
$N=2$ case. It was easily proved that the nonlinear sigma model reduces to a~sine-Gordon model
only over the meridians in the intersection of the coordinate
planes in ${\mathbb R}^3$ with the ${\mathbb S}^2$-sphere, that contain the North and South poles. Consequently, there are sine-Gordon kinks (and multi-solitons) embedded in the nonlinear sigma model living in these meridians. The generalization to any $N$ is immediate:
the $N$ meridians $m_a$, $a=1,\dots, N$, determined by the intersection
of the $\phi_a-\phi_{N+1}$ plane with the ${\mathbb S}^N$-sphere, all of them
containing the points~N and~S,
\[
m_a\Rightarrow \phi_b=0  ,\quad \forall \, b\neq a  , \ \ b=1,\dots, N,
\qquad \phi_a^2+\phi_{N+1}^2=R^2   , \quad a=1,\dots,N   ,
\]
are good trial orbits on which the f\/ield equations (\ref{eqq0}) reduce
to the  system of two PDE's:
\begin{gather}
 \partial_0^2\phi_a -\partial_1^2\phi_a  =  -\alpha_a^2 \phi_a+\lambda  \phi_a  ,\label{eqq1}\\
 \partial_0^2\phi_{N+1} -\partial_1^2\phi_{N+1}  =  -\alpha_{N+1}^2 \phi_{N+1}+\lambda  \phi_{N+1}  .\label{eqq2}
\end{gather}
Use of ``polar'' coordinates in the $\phi_a-\phi_{N+1}$ plane:
\[
\phi_{N+1}=R \cos \theta  ,\qquad \phi_a=\pm R \sin \theta   , \qquad \theta\in [0,\pi]
\]
reveals that the two equations (\ref{eqq1}), (\ref{eqq2}) are tantamount to the sine-Gordon equation in the half-meridians ${\rm sign}  (\phi_a)=\pm 1$:
\begin{gather}
\partial_0^2\theta -\partial_1^2\theta  =  -\tfrac{1}{2} \left( \alpha_a^2-\alpha_{N+1}^2\right)   \sin 2\theta  .\label{sg}
\end{gather}
For the sake of simplicity it is convenient to def\/ine non dimensional parameters and
coordinates in the Minkowski space ${\mathbb
R}^{1,1}$:
\begin{gather*}
\begin{split}
& \sigma_a^2=\frac{\alpha_a^2-\alpha_{N+1}^2}{\alpha_1^2-\alpha_{N+1}^2}, \qquad 1=\sigma_1^2>\sigma_2^2>\dots
>\sigma_N^2>0=\sigma_{N+1}^2,
\\
& t=\sqrt{\alpha_1^2-\alpha_{N+1}^2}    y^0 ,\qquad
x=\sqrt{\alpha_1^2-\alpha_{N+1}^2}    y^1  .
\end{split}
\end{gather*}
Thus, the $N$ sine-Gordon equations (\ref{sg}) become:
\begin{gather}
\Box   \theta=\left( \partial_t^2-\partial_x^2\right)    \theta
=  - \frac{\sigma_a^2}{2}  \sin \ 2\theta     ,\qquad a=1,\dots,N  .\label{sg1}
\end{gather}
The kink/antikink solutions
of the sine-Gordon equations (\ref{sg1}) are the traveling waves
\begin{gather*}
\theta_{K_a}^{\pm}(t,x)  =  2\arctan e^{\pm \sigma_a \left( \gamma
(x-vt)-x_0\right)} , \qquad \gamma=\big(1-v^2\big)^{-\frac{1}{2}} \label{sgtk}
\end{gather*}
that look at rest in their center of mass system $v=0$:
\begin{gather*}
\theta_{K_a}^{\pm}(x)  =  2\arctan e^{\pm \sigma_a (x-x_0)}   .
\end{gather*}
The kinks $\theta_{K_a}^{+}$ belong to the topological
sector ${\cal C}_{\rm NS}$ of the conf\/iguration space, whereas the
anti-kink $\theta_{K_a}^{-}$ lives in  ${\cal C}_{\rm SN}$.
Henceforth, we shall term as topological kinks to these solitary
waves. In the original coordinates in f\/ield space, $\phi_a$ and
$\phi_{N+1}$, $\theta_{K_a}^{\pm}$ correspond to two kinks and their
two anti-kinks, depending of the choice of half-meridian. We will
keep the $K_a$ notation for the $\phi_a>0$ half-meridian and
introduce $K_a^*$ for the $\phi_a<0$ choice:
\begin{gather*}
\phi_a^{K_a}(x) = \frac{R}{\cosh \sigma_a (x-x_0)}  ,\qquad
\phi_{N+1}^{K_a}(x)=\pm R\tanh \sigma_a (x-x_0),\\
\phi_a^{K_a^*}(x) = -\frac{R}{\cosh \sigma_a (x-x_0)}  ,\qquad
\phi_{N+1}^{K_a^*}(x)=\pm R\tanh \sigma_a (x-x_0),\\
\phi_b^{K_a}(x) = \phi_b^{K_a^*}(x)=0   , \qquad \forall \, b\neq a, N+1   .
\end{gather*}
The energy functional, for static conf\/igurations, is
written, in non-dimensional coordinates as:
\begin{gather}
E[\vec{\Phi}]  =  \nu  \int   dx  \left\{ \frac{1}{2}
\frac{d\vec{\Phi}}{dx}\cdot \frac{d\vec{\Phi}}{dx}+\frac{1}{2}
\sum_{a=1}^{N+1} \sigma_a^2   \phi_a^2 -\frac{\alpha_{N+1}^2}{2\nu^2}
 \left(
R^2- \sum_{b=1}^{N+1} \phi_b^2\right)\right\}, \label{energyd}
\end{gather}
where $\nu=\sqrt{\alpha_1^2-\alpha_{N+1}^2}$. The embedded sine-Gordon kink energies are:
\[
E_{K_a}=E_{K_a^*}=2\nu R^2 \sigma_a,  \qquad E_{K_1}> E_{K_2} >\dots >E_{K_N}  .
\]

\begin{figure}[t]
\centerline{\includegraphics[height=4cm]{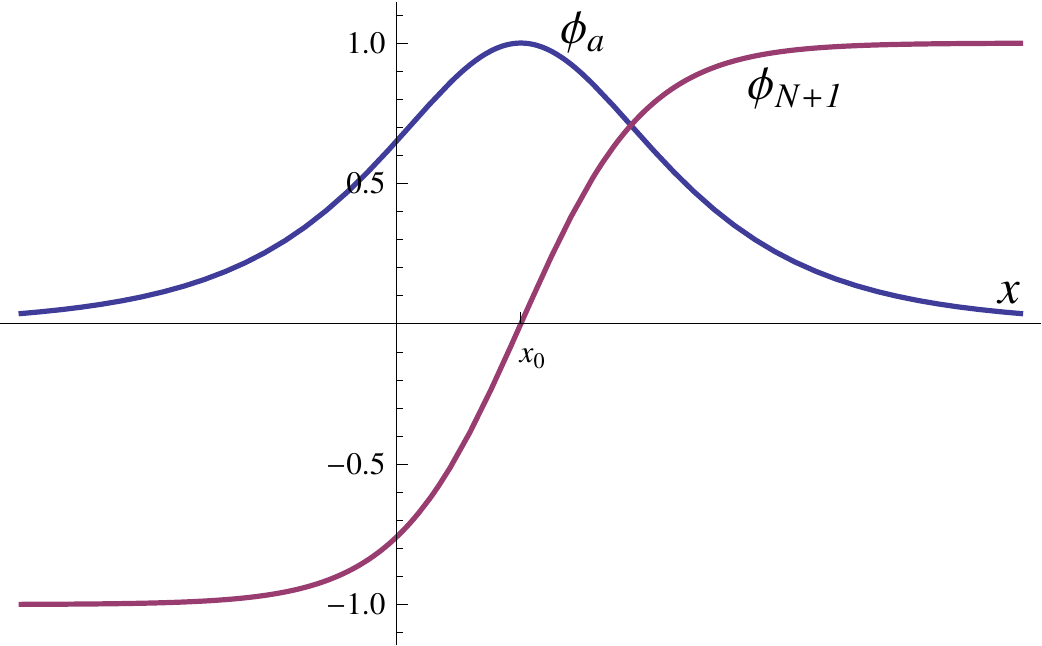}\qquad
\includegraphics[height=4cm]{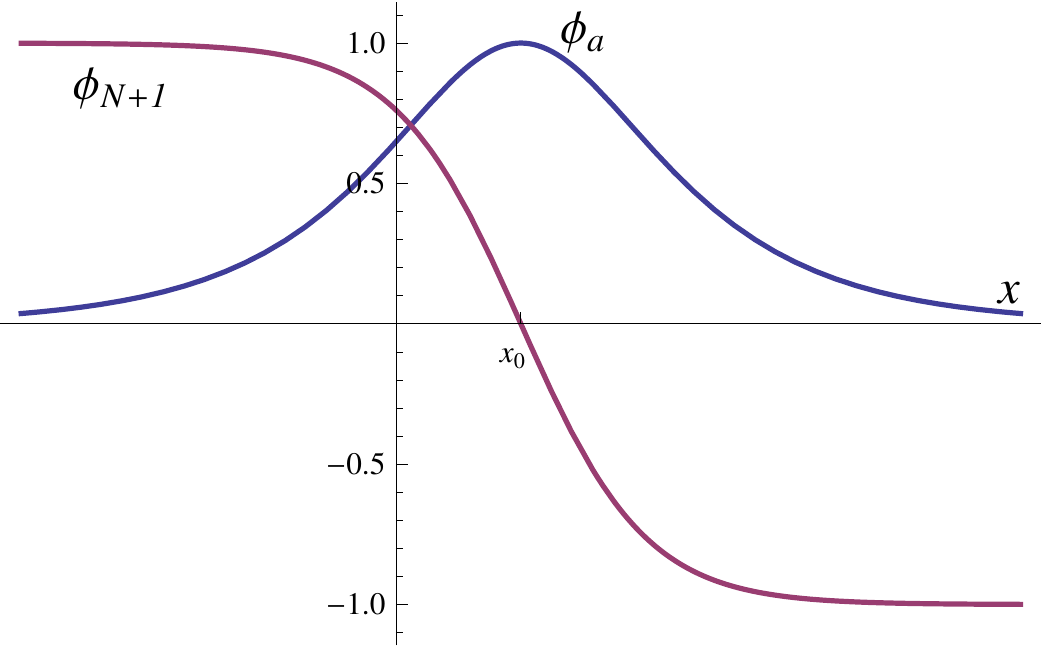}}
\caption{Graphics of $\phi_a^{K_a}(x)$ (blue) and
$\phi_{N+1}^{K_a}(x)$ (red) for kink (left) and antikink (right).}\label{Fig01}
\end{figure}

\section{The moduli space of generic solitary waves}\label{section4}

The system of ordinary dif\/ferential equations (\ref{eqqq}) is the system of motion equations of the repulsive
Neumann problem with $N$ degrees of freedom if one interprets the coordinate $y^1$ as the ``mechanical time''
and the f\/ield components $\phi_a$ as the coordinates setting the ``particle position'', the so-called mechanical analogy~\cite{Rajaraman}. It is a well known
fact that the Neumann system is Hamilton--Jacobi separable using sphero-conical coordinates \cite{Neumann, Moser, Dubrovin}.

Let us introduce sphero-conical coordinates $(u_1,u_2\dots,u_N)$ in ${\mathbb S}^N$ with respect to the set of constants $\bar{\sigma}_a^2  =  1-\sigma_a^2$:
\begin{gather}
\bar{\sigma}_1^2=0<u_1<\bar{\sigma}_2^2<u_2< \cdots<u_N<\bar{\sigma}_{N+1}^2=1   .\label{scc}
\end{gather}
The Cartesian coordinates in ${\mathbb R}^{N+1}$ are given in terms
of the sphero-conical coordinates by:
\begin{gather}
\phi_a^2  =  u_0
\frac{U(\bar{\sigma}_a^2)}{B'(\bar{\sigma}_a^2)} , \qquad
U(z)=\prod_{c=1}^N (z-u_c) ,\qquad B(z)=\prod_{b=1}^{N+1}
\big(z-\bar{\sigma}_b^2\big)  .\label{change}
\end{gather}
The algebraic equation characterizing the
${\mathbb S}^N$-sphere in ${\mathbb R}^{N+1}$ is simply $u_0=R^2$ in
sphero-conical coordinates.

The change of coordinates (\ref{change}) provides a $2^{N+1}$ to $1$
map from the ${\mathbb S}^N$ sphere to the interior of the
hyper-parallelepiped (in the $(u_1,\dots,u_N)$ space) ${\mathbb
P}_N$ characterized by the inequalities (\ref{scc}). The map from
each ``$2^{N+1}$-tant'' of the ${\mathbb S}^N$-sphere to the
interior of ${\mathbb P}_N$ is one-to-one. One ``$2^{N+1}$-tant'' is the piece of ${\mathbb S}^N$ that lives in each
of the $2^{N+1}$ parts in which ${\mathbb R}^{N+1}$ is divided by
the coordinate $N$-hyperplanes. The intersections of the ${\mathbb
S}^N$-sphere with the coordinate $N$-hyperplanes, $N+1$ ${\mathbb
S}^{N-1}$-spheres, are $2^N$ to $1$ mapped into the boundary of
${\mathbb P}_N$, a set of $(N-1)$-hyperplanes in the
$(u_1,\dots,u_N)$ space. Simili modo ${\mathbb S}^{N-k}$-spheres,
intersection of ${\mathbb S}^N$ with coordinate
$(N-k+1)$-hyperplanes of ${\mathbb R}^{N+1}$, are $2^{N-k+1}$ to $1$
mapped into $(N-k)$-hyperplanes of the boundary of ${\mathbb P}_N$.
Finally, the North and South poles of ${\mathbb S}^N$ are mapped
into a single point in the boundary of ${\mathbb P}_N$:
$v^{\pm}\equiv (u_1=0, u_2=\bar{\sigma}_2^2,\dots,
u_N=\bar{\sigma}_N^2)$.

The Euclidean metric in ${\mathbb R}^{N+1}$ $ds^2=d\phi_1^2+\dots+d\phi_{N+1}^2$  becomes in ${\mathbb P}_N$:
\begin{gather}
\left.ds^2\right|_{{\mathbb S}^N}  =  \sum_{a=1}^N g_{aa}
du_a^2 ,\qquad g_{aa}  =  \frac{-R^2}{4}
\frac{U'(u_a)}{B(u_a)}   . \label{metricn}
\end{gather}
Calculations using sphero-conical coordinates remarkably simplify
recalling the Jacobi lemma, see e.g.~\cite{Moser}.
Let be a collection of $p$ distinct constants $(a_1, a_2, \dots  a_p)$, $a_i\neq a_j$, $\forall\, i\neq j$.
Let us def\/ine the function $f(z)=\prod\limits_{i=1}^p (z-a_i)$, $p\geq 2$. The following
identities hold:
\begin{gather}
\sum_{i=1}^p\frac{a_i^k}{f'(a_i)}=0  ,\qquad \forall \, k=0,1,\dots,
p-2  , \qquad \sum_{i=1}^p \frac{a_i^{p-1}}{f'(a_i)}=1  ,\qquad
\sum_{i=1}^p \frac{a_i^{p}}{f'(a_i)}=\sum_{i=1}^p a_i .\label{lemma}
\end{gather}
Now, use of (\ref{lemma}) allows us easily determine the
$V(\vec{\Phi})$ function (\ref{potden}) in sphero-conical
coordinates:
\begin{gather*}
V(u_1,\dots,u_n)  =  \frac{\nu^2 R^2}{2} \left(
1-\sum_{a=1}^{N+1}  \bar{\sigma}_a^2  +  \sum_{b=1}^N
u_b\right).\label{potentialconical}
\end{gather*}
The energy functional for static conf\/igurations (\ref{energyd}) in this system of coordinates
can be written as follows:
\begin{gather}
E[u_1,\dots,u_N]=\nu   \int dx   \left\{ \frac{1}{2} \sum_{a=1}^N
  g_{aa} \left(\frac{du_a}{dx}\right)^2 + \frac{R^2}{2}
\sum_{a=1}^N \frac{\prod\limits_{b=1}^N
(u_a-\bar{\sigma}_b^2)}{U'(u_a)}\right\}.\label{ene}
\end{gather}
The $p-2$ identities on the left in (\ref{lemma}) allows us to write
the potential energy density in dif\/ferent ways. We choose an
expression (\ref{ene}) in the form of a sum of products in the
factors $(u_a-\bar{\sigma}_b^2)$ that guarantees a direct
verif\/ication of the asymptotic conditions (\ref{asy}).

Alternatively, the energy can be written in the Bogomolnyi form \cite{Bogomolny}:
\begin{gather}
E[u_1,\dots,u_n]=\nu  \int   \frac{dx}{2}   \left\{ \sum_{a=1}^N
g_{aa}   \left( \frac{du_a}{dx}-g^{aa} \frac{\partial W}{\partial
u_a} \right)^2\right\} +\nu \int   dx  \sum_{a=1}^N \frac{\partial
W}{\partial u_a}   \frac{du_a}{dx},\label{bogo}
\end{gather}
where $g^{aa}$ denotes the components of the inverse tensor metric,
and $W(u_1,\dots,u_N)$ is a solution of the partial dif\/ferential
equation:
\begin{gather}
\frac{R^2}{2}  \sum_{a=1}^N \frac{\prod\limits_{b=1}^N (u_a-\bar{\sigma}_b^2)}{U'(u_a)}=\frac{1}{2}
  \sum_{a=1}^N   g^{aa} \left( \frac{\partial W}{\partial
u_a}\right)^2 .\label{HJred}
\end{gather}
Note that (\ref{HJred}) is the ``time''-independent
Hamilton--Jacobi equation for zero mechanical energy ($i_1\equiv 0$)
of the ``repulsive'' Neumann problem. Therefore, $W$ is the
zero-energy time-independent characteristic Hamilton function.

The ansatz of separation of variables
$W(u_1,\dots,u_N)=W_1(u_1)+\dots+W_N(u_N)$ reduces the PDE
(\ref{HJred}) to a set of $N$ uncoupled ordinary dif\/ferential
equations depending on $N$ separation constants $i_a$. The
separation process is not trivial (see \cite{Moser}) and requires
the use of equations~(\ref{lemma}). Let write explicitly (\ref{HJred}) as:
\begin{gather}
\sum_{a=1}^N   \frac{1}{U'(u_a)}  \left( \frac{-2 B(u_a)}{R^2}
(W_a'(u_a))^2   -  \frac{R^2}{2}   \prod_{b=1}^N
(u_a-\bar{\sigma}_b^2)  \right)  =  i_1  =  0  .\label{eqqq1}
\end{gather}
The identities of the Jacobi lemma (\ref{lemma}), for $k\leq N-2$, allow to express
(\ref{eqqq1}) as a system of~$N$ equations:
\begin{gather}
\frac{-2 B(u_a)}{R^2}   (W_a'(u_a))^2   -  \frac{R^2}{2}
\prod_{b=1}^N \big(u_a-\bar{\sigma}_b^2\big)    =  i_2   u_a^{N-2}+i_2
u_a^{N-3}+\dots+i_{N-1}   u_a  +  i_N\label{eqqq2}
\end{gather}
depending on $N-1$ arbitrary separation constants $i_2,\dots,i_N$.

Asymptotic conditions (\ref{asy}) imply, in sphero-conical coordinates, the limits: $\lim\limits_{x\to \pm \infty}u_a  =
\bar{\sigma}_a^2$. This requirement f\/ixes, in (\ref{eqqq2}),
all the separation constants $i_a$ to be zero, that correspond to the separatrix trajectories between bounded and unbounded motion in the associated mecha\-ni\-cal system. Thus, integration of (\ref{eqqq2}) gives the $W$ function, i.e.\ the time-independent, zero mechanical energy, characteristic Hamilton function of the repulsive Neumann problem:
\begin{gather}
W(u_1,\dots,u_N)  =  R^2   \sum_{a=1}^N   (-1)^{\delta_a}
\sqrt{1-u_a}      ,     \qquad  \delta_a=0,1   .\label{ww}
\end{gather}
It is interesting to remark that general integration
of the Hamilton--Jacobi equation for the Neumann problem involves
hyper-elliptic theta-functions \cite{Dubrovin}. The asymptotic conditions (\ref{asy})
guaranteeing f\/inite f\/ield theoretical energy to
the solitary waves~-- and f\/inite mechanical action to the associated trajectories~-- force theta functions in the boundary of the Riemann surface where degenerate to very simple irrational functions.

The ordinary dif\/ferential equations:
\begin{gather}
\frac{du_1}{dx}= g^{11}   \frac{\partial W}{\partial u_1}  ,\qquad
\dots ,\qquad  \frac{du_N}{dx}= g^{NN}   \frac{\partial W}{\partial u_N}\label{foe}
\end{gather}
can alternatively be seen as f\/irst-order f\/ield equations for which the f\/ield theoretical energy is of the simple form:
$E[u_1, u_2, \dots, u_N]=\nu \int   dW$, see (\ref{bogo}), or, as the motion equations of the mechanical system restricted
to the sub-space of the phase space such that $i_a=0$, $\forall \, a=1,2, \dots, N$. In dif\/ferential form they look:
\begin{gather}
\frac{dx}{U'(u_a)}=\frac{(-1)^{\delta_a}   du_a}{2  \sqrt{1-u_a}  \prod\limits_{b=1}^n (u_a-\bar{\sigma}_b^2)}
,\qquad a=1,\dots ,N.\label{diffeq}
\end{gather}
Although it is possible to integrate equations (\ref{diffeq}) for an arbitrary value of $N$ (see~\cite{Mumford} for instance), we shall devote the rest of the paper to analyze the explicit solutions and the structure of the moduli
space of solitary waves in the lower $N=2$ and $N=3$ cases.

\section[The moduli space of ${\mathbb S}^2$-solitary waves]{The moduli space of $\boldsymbol{{\mathbb S}^2}$-solitary waves}\label{section5}

\subsection[Embedded sine-Gordon kinks on ${\mathbb S}^2$]{Embedded sine-Gordon kinks on $\boldsymbol{{\mathbb S}^2}$}

\begin{figure}[t]
\centerline{\includegraphics[height=3.5cm]{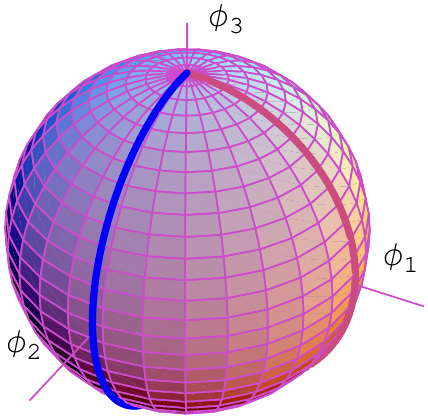}\qquad
\includegraphics[height=4.0cm]{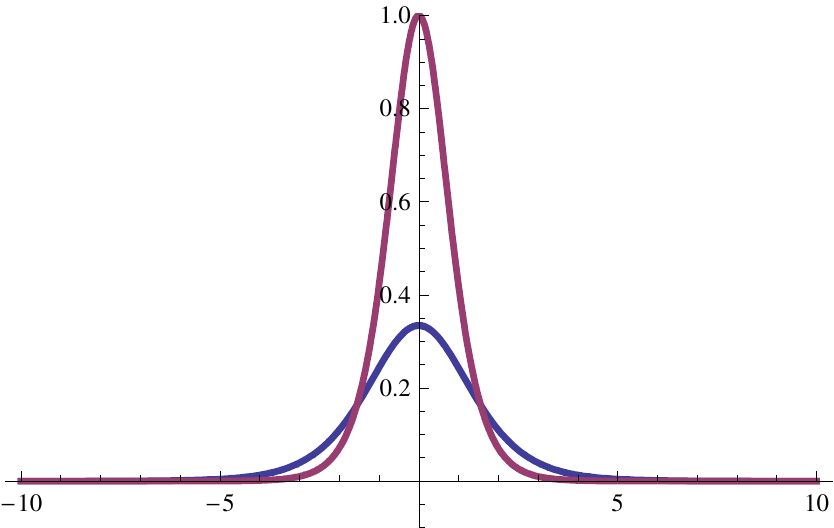}}
\caption{$K_1$ (red) and $K_2$ (blue) kink orbits over ${\mathbb S}^2$ (left).
 Energy densities for $K_1/K_1^*$ (red) and $K_2/K_2^*$
(blue) kinks for a concrete value of $0<\sigma<1$ (right).}\label{Fig02}
\end{figure}

There are two sine-Gordon models embedded in the $\phi_1=0$ and $\phi_2=0$ meridians, see Fig.~\ref{Fig02} (left). The non-dimensional
parameters, for $N=2$, are: $\sigma_1\equiv 1$, $\sigma_2\equiv
\sigma$ and $\sigma_3=0$. The $K_1/K_1^*$ and $K_2/K_2^*$ sine-Gordon kinks/antikinks prof\/iles read:
\begin{gather}
 \phi_1^{K_1}(x)=\frac{R}{\cosh (x-x_0)}  , \qquad
\phi_2^{K_1}(x)=0  ,\qquad \phi_3^{K_1}(x)=\pm R  \tanh (x-x_0),\nonumber \\
  \phi_1^{K_2}(x)=0  , \qquad \phi_2^{K_2}(x)=\frac{R}{\cosh \sigma
(x-x_0)}  ,\qquad \phi_3^{K_2}(x)=\pm R  \tanh \sigma
(x-x_0),\label{k1k2sG}
\\
\phi_1^{K_1^*}(x)=\frac{-R}{\cosh (x-x_0)}  , \qquad
\phi_2^{K_1^*}(x)=0  ,\qquad \phi_3^{K_1^*}(x)=\pm R  \tanh (x-x_0),\nonumber \\
  \phi_1^{K_2^*}(x)=0  , \qquad \phi_2^{K_2^*}(x)=\frac{-R}{\cosh
\sigma (x-x_0)}  ,\qquad \phi_3^{K_2^*}(x)=\pm R  \tanh \sigma
(x-x_0)  .\label{k1k2sG2}
\end{gather}
The $K_1$, $K_1^*$, $K_2$ and $K_2^*$ kinks belong to ${\cal C}_{\rm NS}$, see Fig.~\ref{Fig01}; the $\phi_3$-component is kink shaped for the four kinks, whereas the $\phi_1$- and $\phi_2$-components are bell shaped respectively for~$K_1$ and~$K_2$ but anti-bell shaped for~$K_1^*$ and~$K_2^*$. The antikinks live in ${\cal C}_{\rm SN}$ and have similar prof\/iles. These solitary waves are topological kinks, and they correspond to heteroclinic trajectories in the analogous mechanical system. Their energies are:
\[
E_{K_1}=E_{K_1^*}=2 \nu R^2 ,\qquad
E_{K_2}=E_{K_2^*}=2\nu R^2 \sigma.
\]
The energy of the $K_1/K_1^*$ kinks is greater than the
energy of the $K_2/K_2^*$ kinks, see Fig.~\ref{Fig02} (right).

\subsection[Generic solitary waves on ${\mathbb S}^2$]{Generic solitary waves on $\boldsymbol{{\mathbb S}^2}$}

We shall denote $(\lambda_1,\lambda_2)$ the sphero-conical coordinates in the ${\mathbb S}^2$-sphere, $\lambda_0=R^2$, to avoid confusion with the $N=3$ case. In terms of these coordinates the f\/ields read:
\begin{gather}
\phi_1^2= \frac{R^2}{\bar{\sigma}^2}   \lambda_1
\lambda_2  ,\qquad  \phi_2^2= \frac{R^2}{\sigma^2
\bar{\sigma}^2}   \big(\bar{\sigma}^2-\lambda_1\big)
\big(\lambda_2-\bar{\sigma}^2\big)  ,\qquad \phi_3^2=
\frac{R^2}{\sigma^2}
(1-\lambda_1)(1-\lambda_2)  .\label{change2}
\end{gather}
We recall the def\/inition: $\bar{\sigma}^2=1-\sigma^2$, and the $(\lambda_1,\lambda_2)$ ranges:
\begin{gather}
0  <  \lambda_1 <  \bar{\sigma}^2  <  \lambda_2 <  1  .\label{ineq}
\end{gather}
The change of coordinates (\ref{change2}) maps the ${\mathbb S}^2$-sphere into the
interior of the rectangle ${\mathbb P}_2$ def\/ined by the
inequalities (\ref{ineq}) in the $(\lambda_1,\lambda_2)$-plane. In
fact, the map is eight to one, due to the squares that appears in~(\ref{change2}). Each octant of the ${\mathbb S}^2$-sphere is
mapped one to one into the interior of ${\mathbb P}_2$, whereas the
meridians of ${\mathbb S}^2$ contained in the coordinate planes are
mapped into the boundary of ${\mathbb P}_2$. The set of zeroes ${\cal M}$  of $V(\phi_1,\phi_2,\phi_3)$ in ${\mathbb S}^2$, the North and South poles, becomes a unique point in
${\mathbb P}_2$: $ v^{\pm}\equiv (\phi_1,\phi_2,\phi_3)=(0,0,\pm R)
\Rightarrow v^{\pm}\equiv (\lambda_1,\lambda_2)=(0,\bar{\sigma}^2)$.
The quadratic polynomial $V(\phi_1,\phi_2,\phi_3)$ becomes linear in sphero-conical coordinates:
\begin{gather*}
V(\lambda_1,\lambda_2)=  \frac{\nu^2 R^2}{2} \left(
\lambda_1+\lambda_2 -\bar{\sigma}^2\right)
\end{gather*}
and the energy functional for static conf\/igurations in terms of $(\lambda_1,\lambda_2)$ is:
\begin{gather}
E[\lambda_1,\lambda_2]=\nu  \int dx   \left\{ \frac{1}{2}
 \sum_{i=1}^2 g_{ii} \left(\frac{d\lambda_i}{dx}\right)^2+
\frac{R^2}{2}  \left( \frac{\lambda_1
(\lambda_1-\bar{\sigma}^2)}{\lambda_1-\lambda_2}+\frac{\lambda_2
(\lambda_2-\bar{\sigma}^2)}{\lambda_2-\lambda_1}\right)\right\}  .\label{eee}
\end{gather}
Here, the components of the metric tensor (\ref{metricn}) read:
\[
g_{11}=\frac{-R^2(\lambda_1-\lambda_2)}{4\lambda_1
 (\bar{\sigma}^2-\lambda_1)(1-\lambda_1)}  ,\qquad
g_{22}=\frac{-R^2(\lambda_2-\lambda_1)}{4\lambda_2
 (\bar{\sigma}^2-\lambda_2)(1-\lambda_2)}  .
\]
The knowledge of the solution (\ref{ww}) for $N=2$:
\begin{gather*}
W(\lambda_1,\lambda_2)=R^2   \left(   (-1)^{\delta_1}
\sqrt{1-\lambda_1}   +  (-1)^{\delta_2}  \sqrt{1-\lambda_2}  \right)  ,\qquad \delta_1,\delta_2=0,1  ,
\end{gather*}
suggests to write the energy functional (\ref{eee}) in the Bogomolnyi form (\ref{bogo}) leading to the system of f\/irst-order ODE's (\ref{foe}):
\begin{gather*}
\frac{d\lambda_1}{dx} = (-1)^{\delta_1}   \frac{2\lambda_1
(\lambda_1-\bar{\sigma}^2) \sqrt{1-\lambda_1}}{
 (\lambda_1-\lambda_2)}  ,\qquad  \frac{d\lambda_2}{dx} = (-1)^{\delta_2}   \frac{2\lambda_2
(\lambda_2-\bar{\sigma}^2) \sqrt{1-\lambda_2}}{
 (\lambda_2-\lambda_1)}  .
 \end{gather*}
These equations can be separated in a easy way:
\begin{gather}
0 = \frac{d\lambda_1}{(-1)^{\delta_1}   2\lambda_1  (\lambda_1-\bar{\sigma}^2)
\sqrt{1-\lambda_1}} +\frac{d\lambda_2}{(-1)^{\delta_2}   2\lambda_2
(\lambda_2-\bar{\sigma}^2) \sqrt{1-\lambda_2}},\label{dd2a}\\
dx = \frac{d\lambda_1}{(-1)^{\delta_1}   2  (\lambda_1-\bar{\sigma}^2)
\sqrt{1-\lambda_1}} +\frac{d\lambda_2}{(-1)^{\delta_2}   2
(\lambda_2-\bar{\sigma}^2) \sqrt{1-\lambda_2}}  .\label{dd2b}
\end{gather}
The Hamilton--Jacobi procedure precisely prescribes the equation (\ref{dd2a})
as the rule satisf\/ied by the orbits of zero mechanical energy whereas (\ref{dd2b})
sets the mechanical time schedule of these separatrix trajectories between bounded
and unbounded motion.

Instead of attempting a direct solution of the system (\ref{dd2a}), (\ref{dd2b}) we introduce
new variables $s_1=(-1)^{\delta_1}\sqrt{1-\lambda_1}$, $s_2=(-1)^{\delta_2}\sqrt{1-\lambda_2}$.
In terms of $s_1$ and $s_2$ the ODE's system (\ref{dd2a}), (\ref{dd2b}), after decomposition in
simple fractions, becomes:
\begin{gather}
\sum_{a=1}^2  \frac{ds_a}{1-s_a^2}=dx     ,  \qquad \sum_{a=1}^2  \frac{ds_a}{\sigma^2-s_a^2}=dx  . \label{dd2}
\end{gather}
Integrating equations (\ref{dd2}) in terms of the inverse of
hyperbolic tangents/cotangents, and using the addition formulas for these
functions we f\/ind the following general solution depending on two real integration
constants $\gamma_1$ and $\gamma_2$:
\begin{gather}
\frac{s_1s_2}{1+s_1s_2}=t_1   ,\qquad \frac{\sigma^2+s_1s_2}{\sigma (s_1+s_2)}=t_2   , \qquad t_1=\tanh(x+\gamma_1)  ,   \qquad t_2=\tanh \sigma (x+\gamma_2)  .\label{viet}
\end{gather}
To solve the system of equations (\ref{viet}) separately in $s_1$ and $s_2$  we introduce
the new variables $A=s_1+s_2$ and $B=s_1s_2$. The subsequent linear system in $A$ and $B$ and its
solution are:
\begin{gather*}
A-t_1B=t_1       ,   \qquad    \sigma t_2A-B=\sigma^2, \qquad A=\frac{(1-\sigma^2)t_1}{1-\sigma t_1t_2}
      , \qquad      B=\frac{\sigma t_1t_2-\sigma^2}{1-\sigma t_1t_2}   .
\end{gather*}
Therefore, $s_1$ and $s_2$ are the roots of the quadratic equation
\begin{gather*}
s^2  -  A  s  +  B    = 0   , \qquad s_1=\frac{A+\sqrt{A^2-4 B}}{2}       ,   \qquad    s_2=\frac{A-\sqrt{A^2-4 B}}{2}  .
\end{gather*}
Because $\lambda_1=1-s_1^2$, $\lambda_2=1-s_2^2$, the come back to Cartesian coordinates (\ref{change2}) provides the explicit analytical formulas for the two-parametric family of solitary waves:
\begin{gather}
\phi_1(x) =  (-1)^{\epsilon_1}   R\bar{\sigma}   \frac{\sech
(x+\gamma_1)}{1-\sigma
\tanh (x+\gamma_1) \tanh \sigma (x+\gamma_2)}   ,\qquad   \epsilon_1=0,1,\nonumber\\
\phi_2(x) =  (-1)^{\epsilon_2}  R \bar{\sigma}   \frac{\tanh
(x+\gamma_1) \sech \sigma (x+\gamma_2)}{1-\sigma \tanh
(x+\gamma_1) \tanh \sigma (x+\gamma_2)}   ,\qquad \epsilon_2=0,1,\label{sol2}\\
\phi_3(x) =
(-1)^{\epsilon_3}  R   \frac{\sigma-\tanh (x+\gamma_1)\tanh \sigma
(x+\gamma_2)}{1-\sigma \tanh (x+\gamma_1) \tanh \sigma
(x+\gamma_2)}   ,    \qquad   \epsilon_3=0,1  .\nonumber
\end{gather}

\subsection{The structure of the moduli space of kinks}

\begin{figure}[t]
\centerline{\includegraphics[height=3.2cm]{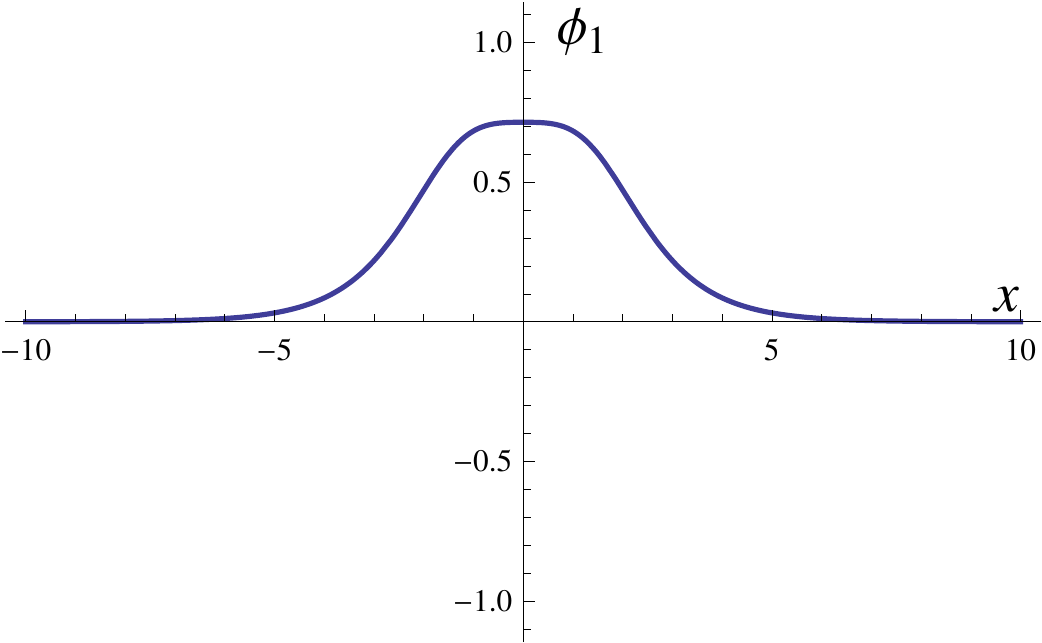}\
\includegraphics[height=3.2cm]{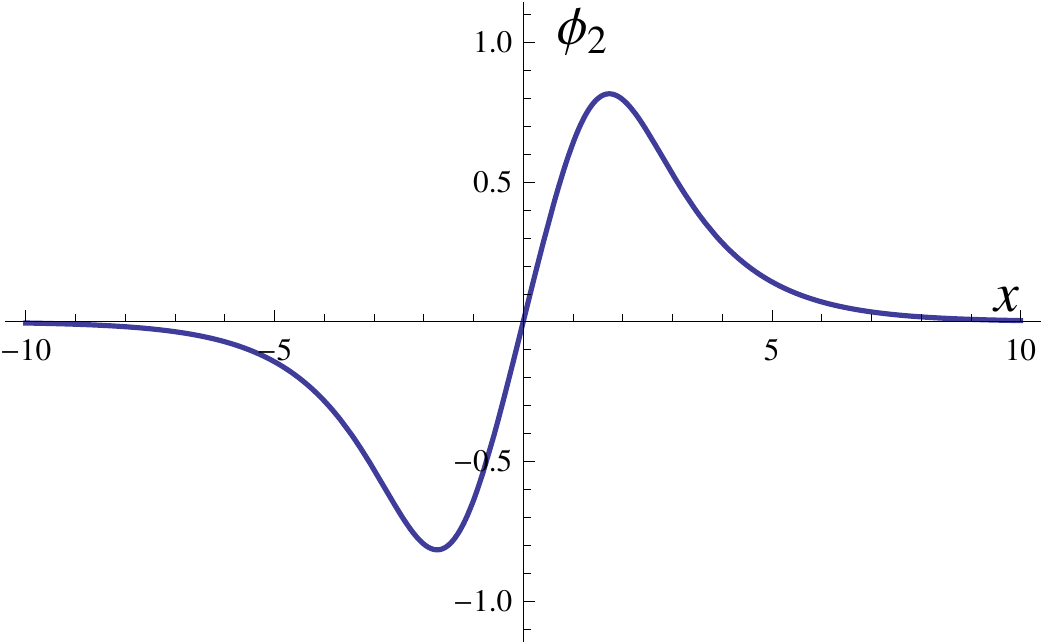}\  \includegraphics[height=3.2cm]{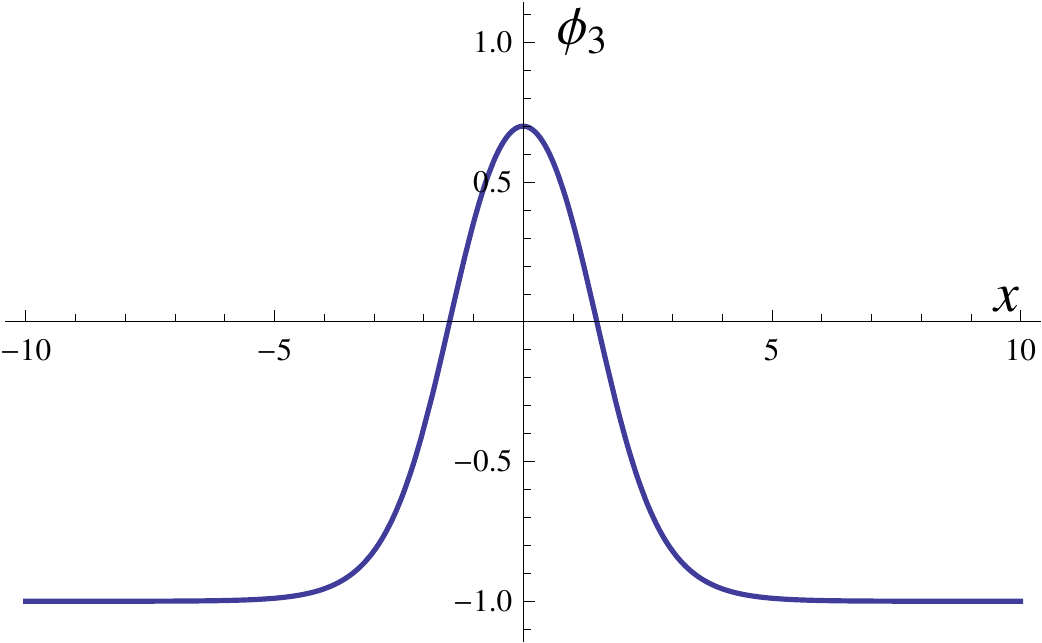}}
\caption{Graphics of (\ref{sol2}) corresponding to the following values of parameters: $R=1$,
$\sigma=0.7$, $\gamma=0$, $\bar{\gamma}=0$, i.e.\ the maximally
symmetric generic kink (it is chosen
$\epsilon_1=\epsilon_2=\epsilon_3=0$).}\label{Fig03}
\end{figure}

\begin{figure}[t]
\centerline{\includegraphics[height=3.2cm]{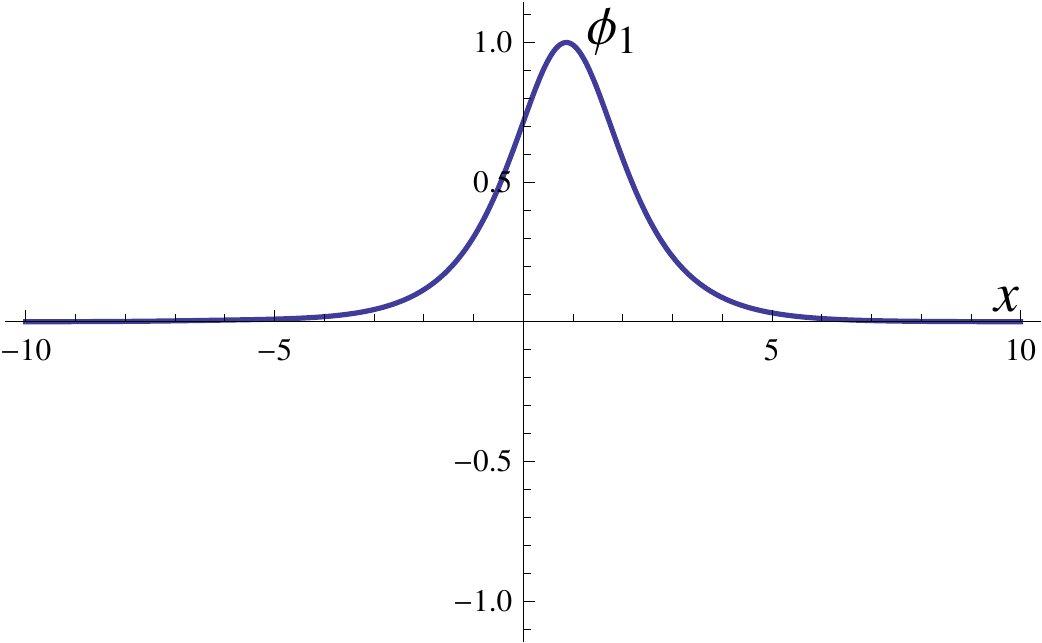}\
\includegraphics[height=3.2cm]{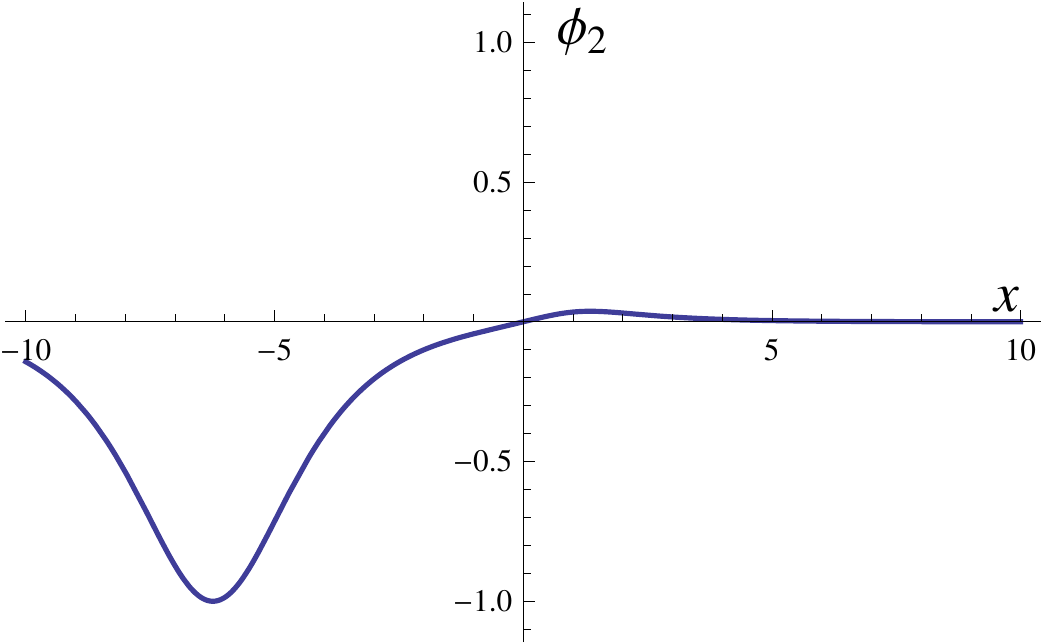}\  \includegraphics[height=3.2cm]{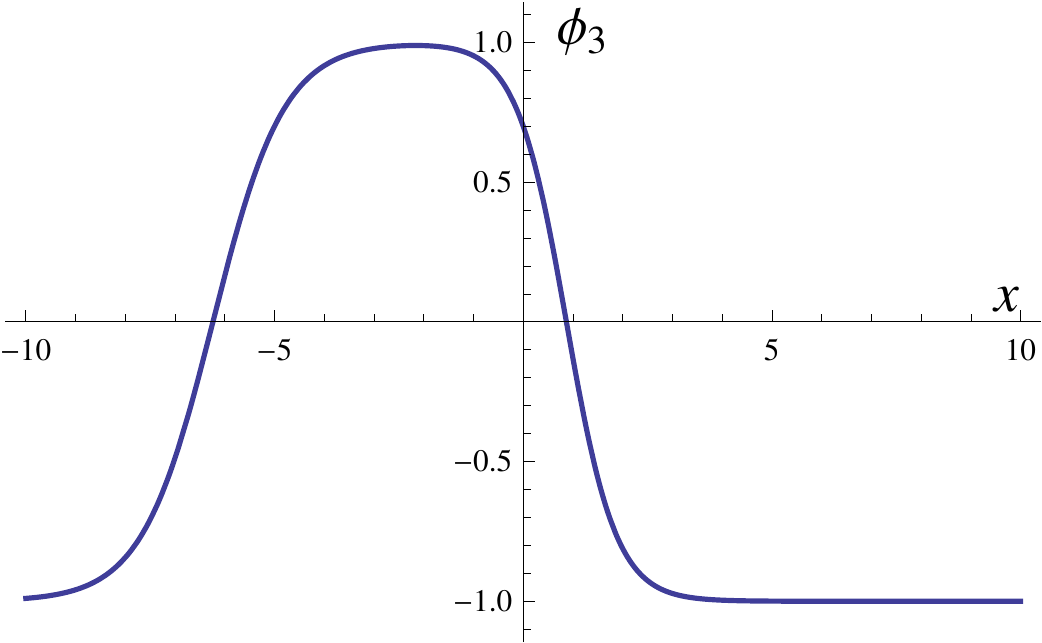}}
\caption{Graphics of (\ref{sol2}) corresponding to the following values of parameters: $R=1$,
$\sigma=0.7$, $\gamma=0$, $\bar{\gamma}=5$,
 $\epsilon_1=\epsilon_2=\epsilon_3=0$.}\label{Fig04}
\end{figure}

\begin{figure}[t!]
\centerline{\includegraphics[height=3.5cm]{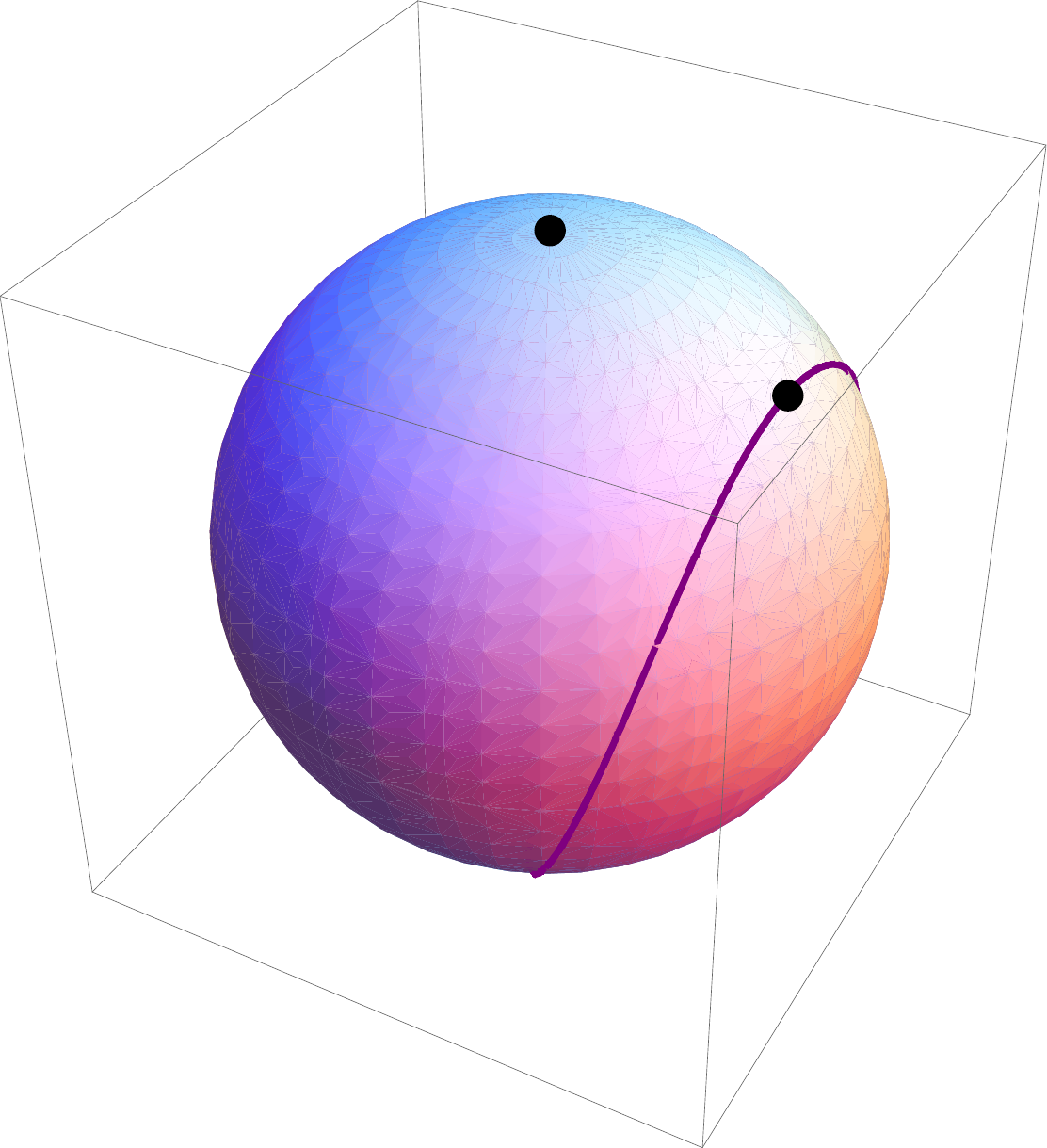}\qquad
\includegraphics[height=3.5cm]{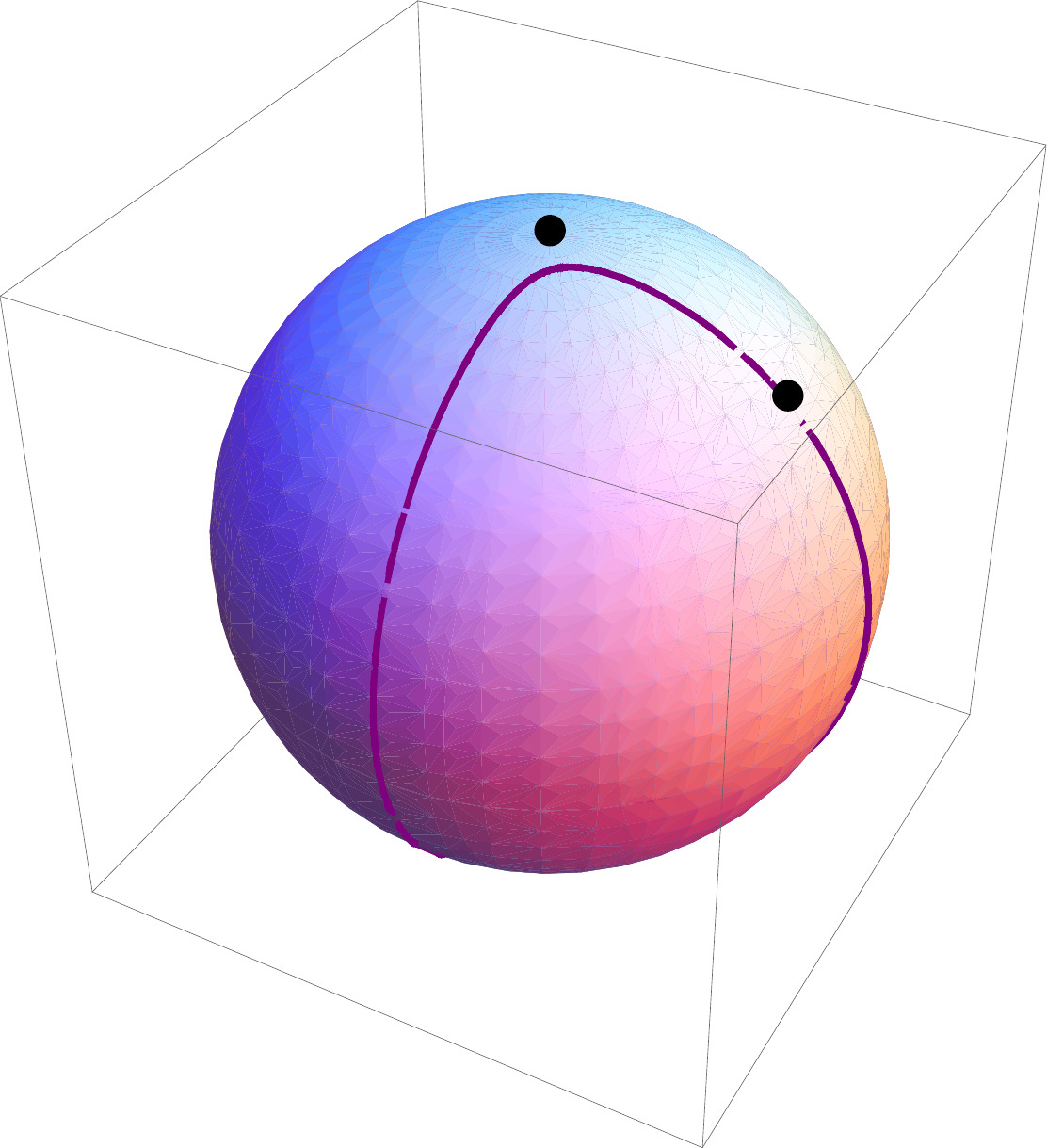}\qquad \includegraphics[height=3.5cm]{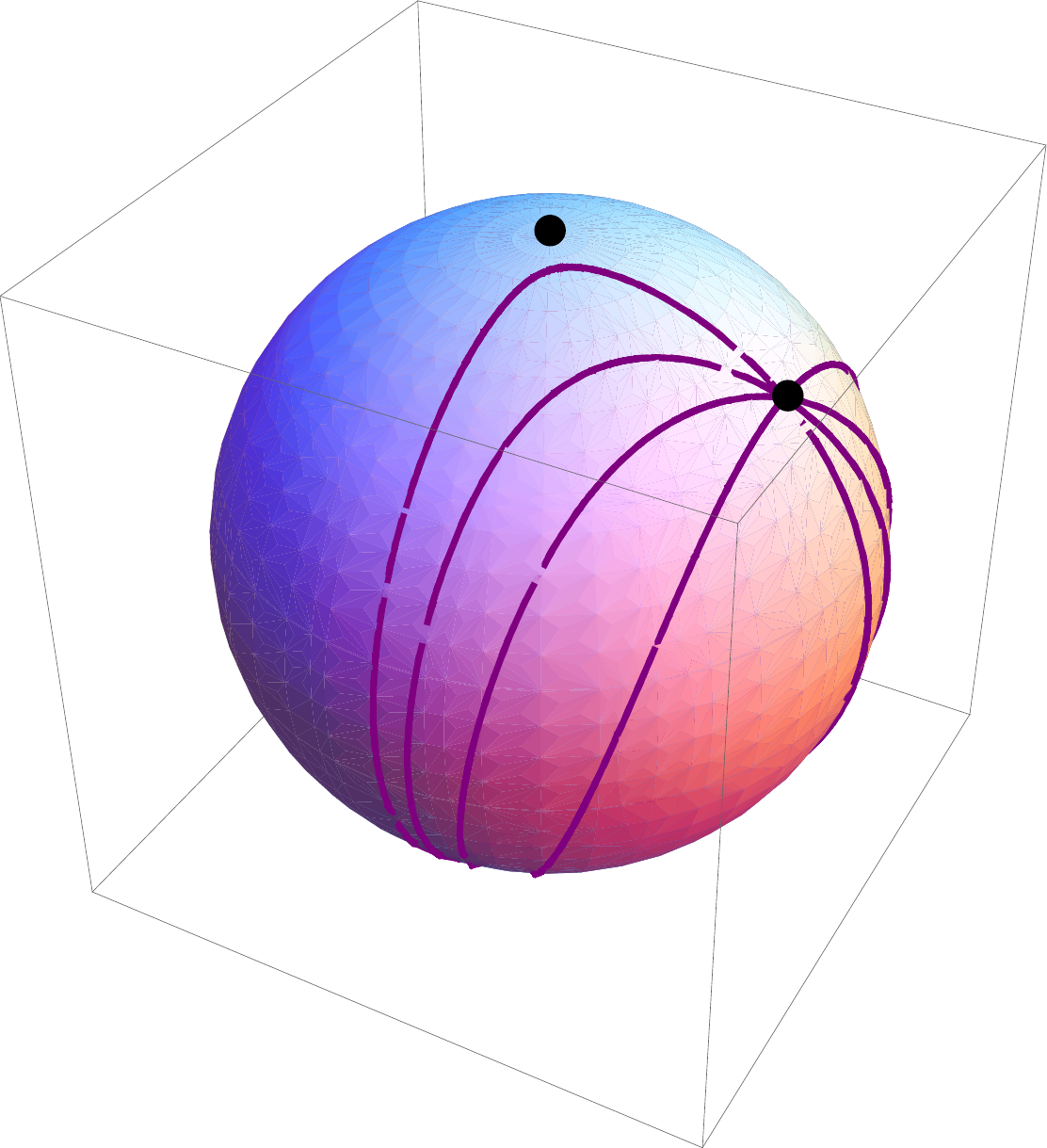}}
\caption{$NTK$ kink-orbits, corresponding to: $R=1$, $\sigma=0.7$,
$\epsilon_1=\epsilon_2=\epsilon_3=0$ and  $\bar{\gamma}=0$ (left),
 $\bar{\gamma}=5$ (middle).  Several kink orbits in ${\cal
C}_{\rm SS}$ corresponding to several dif\/ferent values of
$\bar{\gamma}$ (right).}\label{Fig05}
\end{figure}

\begin{figure}[t]
\centerline{\includegraphics[height=3.5cm]{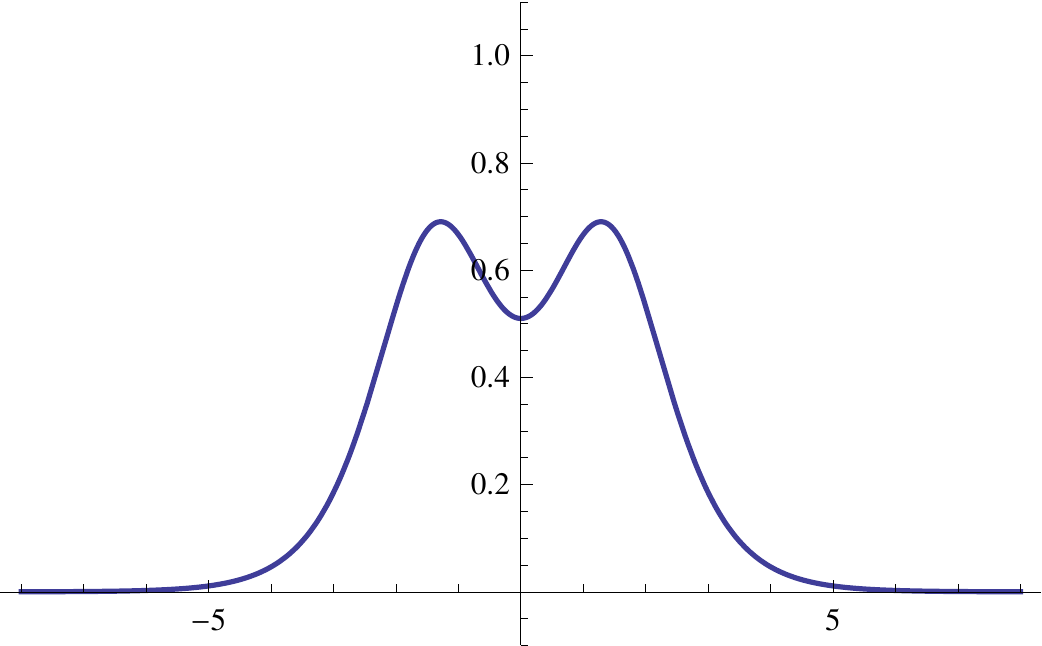}\quad
\includegraphics[height=4cm]{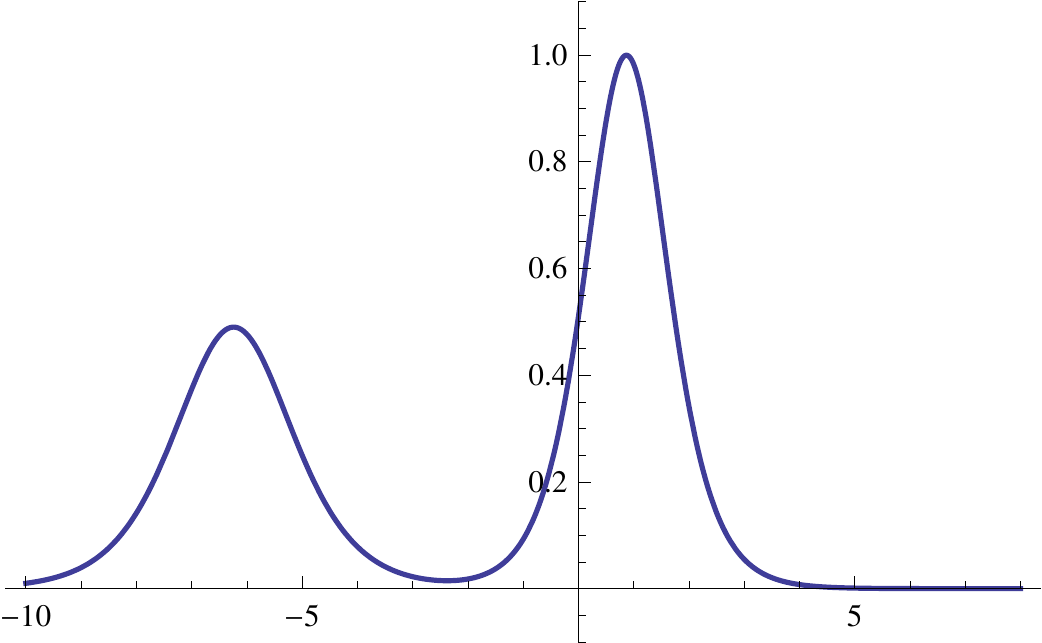}} \caption{Kink energy
density for the $NTK$ determined by the constants: $R=1$, $\sigma=0.7$, and   $\gamma=\bar{\gamma}=0$ (left),
 $\gamma=0$ and $\bar{\gamma}=5$ (right).}\label{Fig06}
\end{figure}

To describe the structure of the two-dimensional moduli space of
solitary waves solutions (\ref{sol2}) it is convenient to rely on a
re-shuf\/f\/ling of the coordinates of the moduli space:
$\gamma=-\gamma_1$, $\bar{\gamma}=\gamma_2-\gamma_1$. $\gamma$
determines the ``center'' of the energy density of a given kink and
$\bar{\gamma}$ distinguishes between dif\/ferent kinks in the moduli
space by selecting the kink orbit in ${\mathbb S}^2$, see Figs.~\ref{Fig05}
and~\ref{Fig06}. In the Figs.~\ref{Fig03} and~\ref{Fig04}, however, we have plotted the
components of the kink prof\/iles for two dif\/ferent choices of
$\bar{\gamma}$ and $\gamma=0$. We remark that none of the three
components are kink-shaped, the $\phi_1$ and $\phi_3$ components are
bell-shaped and the $\phi_2$-components have a~maximum and a~minimum. The dif\/ference is that the $x\to -x$ ref\/lection symmetry is
lost for $\bar{\gamma}=5$. The prof\/iles of these two solitary waves
tend to the South pole in both $x\to\pm\infty$. Therefore, these
kinks are non-topological kinks living in ${\cal C}_{\rm SS}$.

 ($\gamma$, $\bar{\gamma}$) are good coordinates in the
solitary wave moduli space characterized by the orbits in~${\mathbb
P}_2$. The inverse mapping to ${\mathbb S}^2$ is classif\/ied by the
dif\/ferent choices of $\epsilon_1$, $\epsilon_2$ and $\epsilon_3$:
\begin{itemize}\itemsep=0pt

\item The value of $\epsilon_3$ selects the
topological sector of the solution. Clearly:
\[
\epsilon_3=0\Rightarrow \lim_{x\to \pm \infty}\phi_3  =
-R, \qquad \epsilon_3=1\Rightarrow \lim_{x\to \pm
\infty}\phi_3  =  R
\]
and the equations (\ref{sol2}) determine two families of
non-topological ($NTK$) kinks, one family living in ${\cal C}_{\rm
SS}$, $\epsilon_3=0$, and the other one belonging to ${\cal C}_{\rm
NN}$, $\epsilon_3=1$. They are homoclinic trajectories of the analogous mechanical system.

\item $\epsilon_1=0$ provides the $NTK$ kink orbits
running in the $\phi_1>0$ hemisphere and $\epsilon_1=1$ corresponds
to the $NTK$ orbits passing through the other ($\phi_1<0$)
hemisphere of ${\mathbb S}^2$.

\item $\epsilon_2$, however, does not af\/fect to the
kink orbit and merely specif\/ies the kink/antikink cha\-rac\-ter of the
solitary wave, i.e.\ the sense in which the orbit is traveled.
\end{itemize}

  In Fig.~\ref{Fig05} (right) several $NTK$ orbits are shown in
${\mathbb S}^2$. The sector ${\cal C}_{\rm SS}$ and the $\phi_1>0$
hemisphere have been selected. All of them cross the $\phi_2=0$
meridian at the same point: $F\equiv (R\bar{\sigma},0,R\sigma)$ -- in
sphero-conical coordinates this point is the low right corner
$\lambda_1=\lambda_2=\bar{\sigma}^2$ of ${\mathbb P}^2$). Therefore
the point $F$ is a conjugate point{\footnote{We shall show in
Section~\ref{section7} that there exists a Jacobi f\/ield orthogonal to each $NTK$
orbit that becomes zero at~S and~$F$, conf\/irming thus that they
are conjugate points. We shall also prof\/it of this fact to show the
instability of the $NTK$ kinks by means of the Morse index theorem.}}
of the South pole S of ${\mathbb S}^2$.

In Fig.~\ref{Fig06} the graphics of the kink energy densities of
two members of the $NTK$ family are plotted. We remark that the energy
density appears to be composed by two basic elements.
This behavior is more evident for $|\bar{\gamma}| \gg 0$ because in such a range the energy density resembles the superposition of the energy densities of the two embedded sine-Gordon kinks (compare with Fig.~\ref{Fig02} (right)). All the $NTK$ kinks, however, have the same energy that can be computed from the formula (\ref{bogo}):
\begin{gather*}
E_{NTK} = 2\nu R^2(1+\sigma)  =  E_{K_1}+E_{K_2}  .\label{ksr}
\end{gather*}
There is a kink energy sum rule: the energy of any $NTK$ kink is exactly the
the sum of the two sine-Gordon, $K_1$ and $K_2$, topological kink energies. These sum rules
arises in every f\/ield theoretical model with mechanical analogue system which is
Hamilton--Jacobi separable and has several unstable equilibrium points, see, e.g.,~\cite{Nonlinearity}. Moreover, looking at Fig.~\ref{Fig04} as a posterior picture to Fig.~\ref{Fig03} in a movie that evolves with increasing $\bar{\gamma}$,
it is clear that we f\/ind almost a $K_1$ anti-kink in the $x\gg  0$ region and
a $K_2^*$ kink in the $x\ll 0$ region when $\bar{\gamma}\to\infty$. This means that
the solitary waves (\ref{sol2}) tend to a $K_1$ kink/antikink plus a
$K_2$ antikink/kink when $\bar{\gamma}\to\pm\infty$. Thus the combinations $K_1+K_2$ (kink/antikink and
antikink/kink) form the closure of the $NTK$
moduli space and belong to the boundary. Alternatively, one can tell that
there exist only two basic (topological) kink solutions and the rest of the kinks, the $NTK$ family, are (nonlinear) combinations of the two basic solitary waves.

\section[The moduli space of ${\mathbb S}^3$-solitary waves]{The moduli space of $\boldsymbol{{\mathbb S}^3}$-solitary waves}\label{section6}

\subsection[Embedded $N=2$ kinks]{Embedded $\boldsymbol{N=2}$ kinks}

In the massive nonlinear ${\mathbb S}^3$-sigma model we f\/irst count the three embedded sine-Gordon kinks:
\begin{itemize}\itemsep=0pt
\item The $K_1/K_1^*$ kinks and their anti-kinks living in the meridian
\[
\left\{\phi_1^2+\phi_4^2=R^2 \right\}      \equiv       \left\{ \{ \phi_2=\phi_3=0 \}    \bigcap     \left\{ \phi_1^2+\phi_2^2+\phi_3^2+\phi_4^2=R^2\right\} \right\}    ,
\]
the intersection of the $\phi_2=\phi_3=0$ plane with the ${\mathbb S}^3$ sphere.

\item Idem for the $K_2/K_2^*$ and their anti-kinks:
\[
 \left\{\phi_2^2+\phi_4^2=R^2 \right\}      \equiv       \left\{ \{\phi_1=\phi_3=0\}     \bigcap     \left\{ \phi_1^2+\phi_2^2+\phi_3^2+\phi_4^2=R^2\right\} \right\}    .
 \]

\item Idem for the $K_3/K_3^*$ kinks and their anti-kinks:
\[
 \left\{ \phi_3^2+\phi_4^2=R^2\right\}      \equiv       \left\{\{ \phi_1=\phi_2=0 \}    \bigcap    \left\{ \phi_1^2+\phi_2^2+\phi_3^2+\phi_4^2=R^2\right\} \right\}    .
 \]
\end{itemize}
The novelty is that all the non-topological kinks of the massive nonlinear ${\mathbb S}^2$-sigma model are embedded in the ${\mathbb S}^3$ version in three dif\/ferent ${\mathbb S}^2$ ``meridian'' sub-manifolds.

\begin{enumerate}\itemsep=0pt
\item Consider the two-dimensional ${\mathbb S}^2_I$ sphere:
\[
\left\{ \phi_2^2+\phi_3^2+\phi_4^2=R^2 \right\}       \equiv       \left\{ \{\phi_1=0\}    \bigcap
\left\{\phi_1^2+\phi_2^2+\phi_3^2+\phi_4^2=R^2\right\} \right\}       ,
\]
the intersection of the $\phi_1=0$ 3-hyperplane in ${\mathbb R}^4$ with the
${\mathbb S}^3$-sphere.

The restriction of the $N=3$-model to these two-manifold collects all the solitary waves of the $N=2$-model.
We shall call $NTK_I$ kinks to the $N=2$ $NTK$ kinks of the $N=2$ model that lives in ${\mathbb S}^2_I$.
The $K_2/K_2^*$ and $K_3/K_3^*$ sine-Gordon
topological kinks/anti-kinks also live in in ${\mathbb S}^2_I$ and belong to the boundary
of the $NTK_I$ moduli space.

\item Idem for the ${\mathbb S}^2_{II}$ sphere:
\[
\left\{ \phi_1^2+\phi_3^2+\phi_4^2=R^2 \right\}       \equiv       \left\{ \{\phi_2=0\}    \bigcap
\left\{\phi_1^2+\phi_2^2+\phi_3^2+\phi_4^2=R^2\right\} \right\}       .
\]
The non-topological kinks running on this 2-sphere will be termed $NTK_{II}$. The
sine-Gordon kinks/antikinks are: $K_1/K_1^*$ and $K_3/K_3^*$.

\item Idem for the ${\mathbb S}^2_{III}$ sphere:
\[
\left\{ \phi_1^2+\phi_2^2+\phi_4^2=R^2 \right\}       \equiv       \left\{\{\phi_3=0\}    \bigcap
\left\{\phi_1^2+\phi_2^2+\phi_3^2+\phi_4^2=R^2\right\} \right\}    .
\]
The non-topological kinks running on this 2-sphere will be termed $NTK_{III}$. The
sine-Gordon kinks/antikinks are: $K_1/K_1^*$ and $K_2/K_2^*$.
\end{enumerate}
In the ``equatorial'' sphere $\phi_1^2+\phi_2^2+\phi_3^2=R^2$, $\phi_4=0$, of ${\mathbb S}^3$, however, there are no kinks because the equilibrium points, the North and South poles, are not included.

\subsection{Generic solitary waves}

\begin{figure}[t]
\centerline{\includegraphics[height=3.5cm]{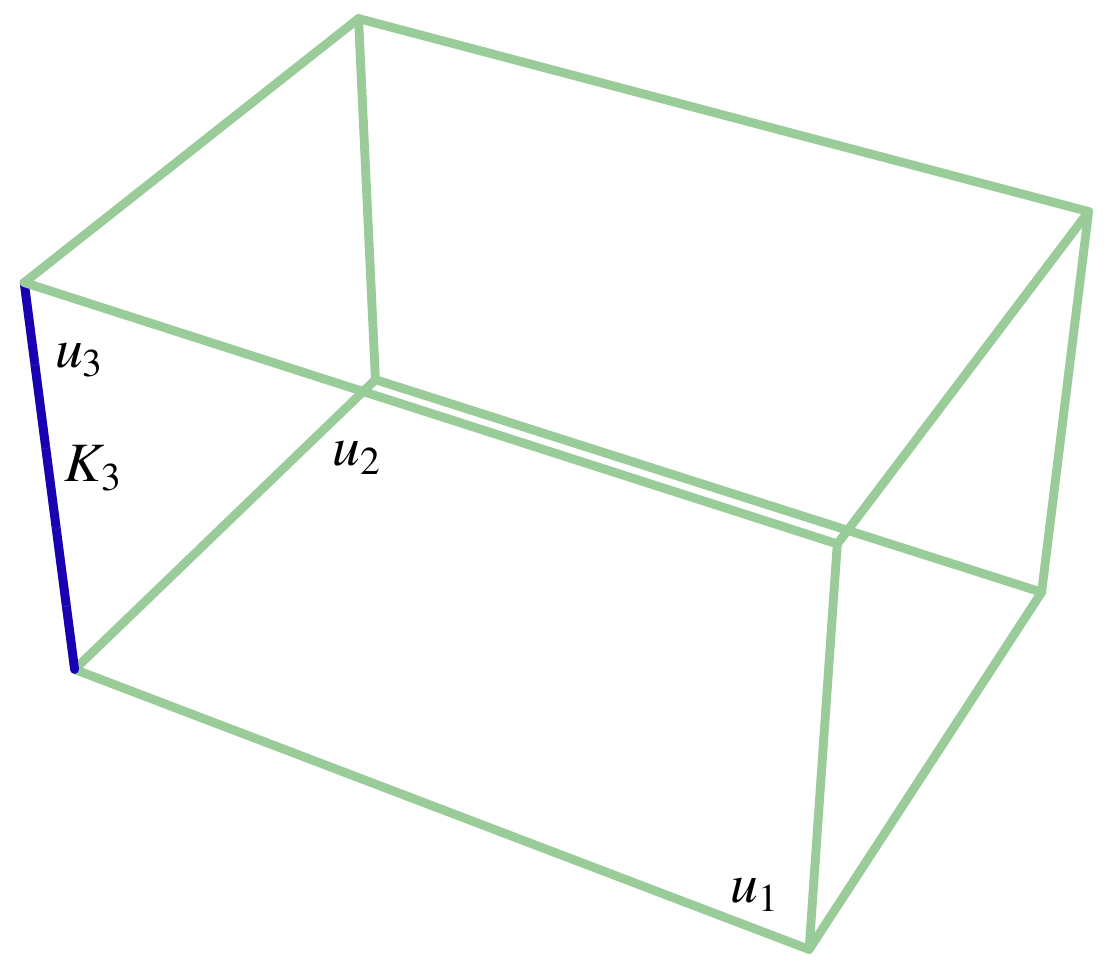}\quad
\includegraphics[height=3.5cm]{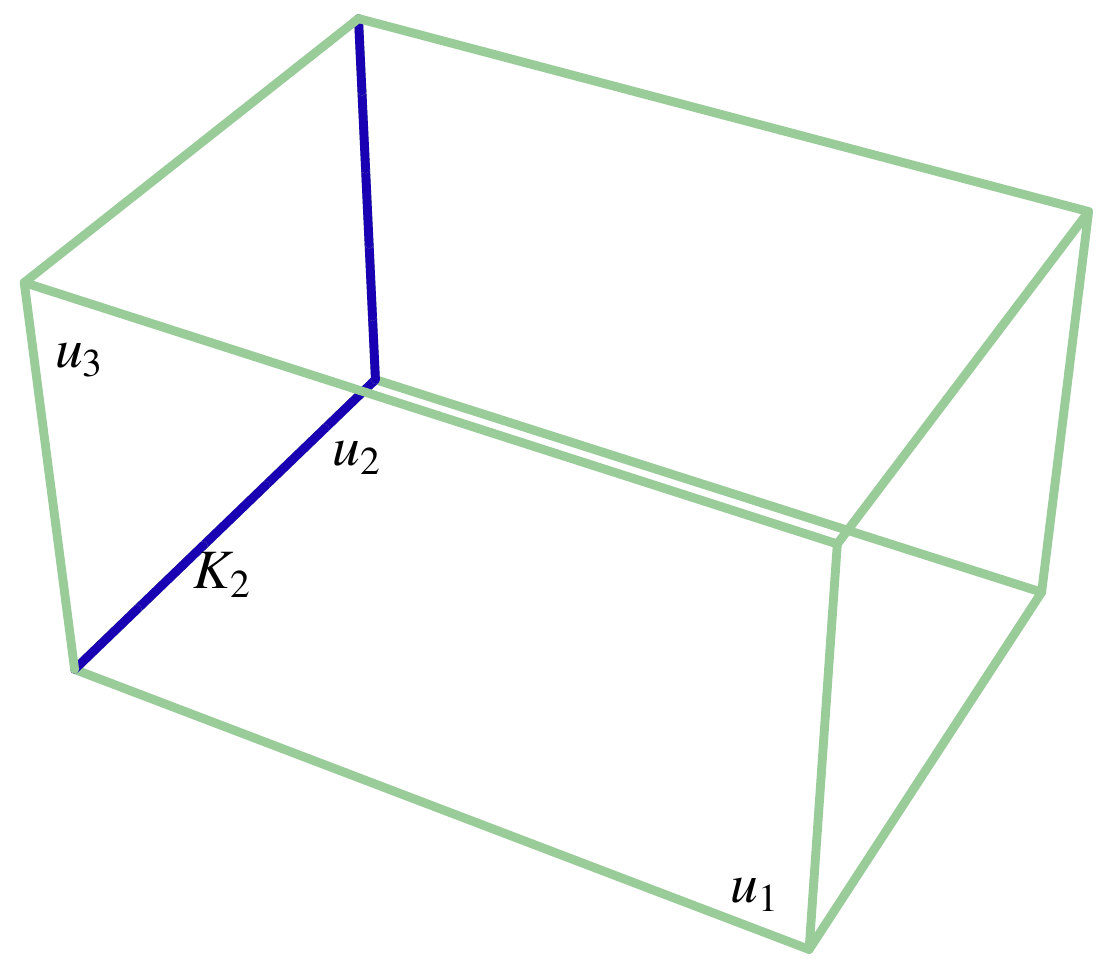}\quad \includegraphics[height=3.5cm]{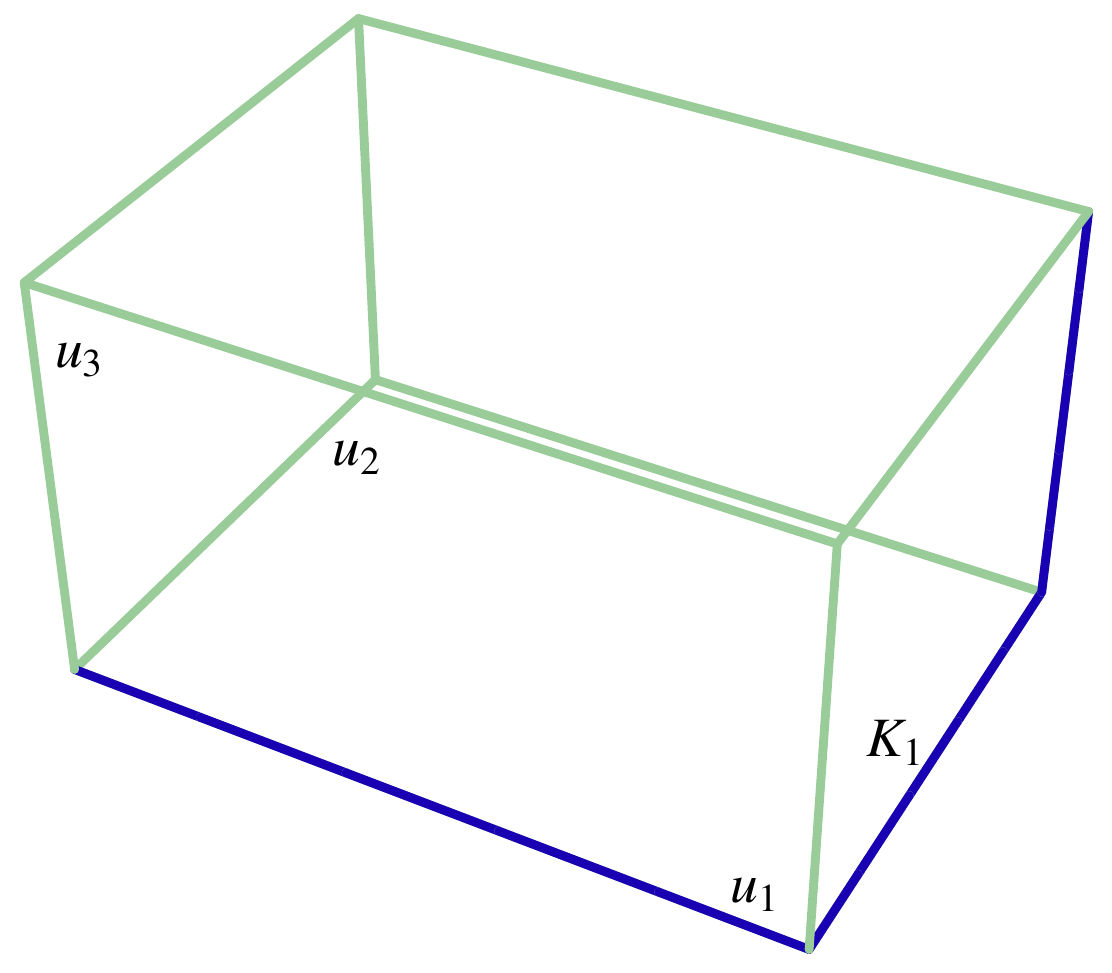}} \caption{Parallelepiped ${\mathbb P}_3$. The $K_1$, $K_2$ and $K_3$ sine-Gordon kink orbits
are depicted (in blue) at the corresponding edges of ${\mathbb P}_3$.}\label{Fig07}
\end{figure}

Away from the meridian 2-spheres ${\mathbb S}^2_I$, ${\mathbb S}^2_{II}$ and ${\mathbb S}^2_{III}$ considered in the previous sub-section, there is a three-parametric family of generic kinks. In $N=3$ dimensional sphero-conical coordinates~(\ref{change}):
\begin{gather}
\phi_1^2 =  u_0
\frac{u_1 u_2 u_3}{\bar{\sigma}_2^2\bar{\sigma}_3^2}  , \qquad \phi_2^2= u_0  \frac{(\bar{\sigma}_2^2-u_1)
(\bar{\sigma}_2^2-u_2) (\bar{\sigma}_2^2-u_3)}{\sigma_2^2 \bar{\sigma}_2^2 (\bar{\sigma}_3^2-\bar{\sigma}_2^2)}  ,\nonumber \\
\phi_3^2 =  u_0  \frac{(\bar{\sigma}_3^2-u_1)(\bar{\sigma}_3^2-u_2)
(\bar{\sigma}_3^2-u_3)}{\sigma_3^2\bar{\sigma}_3^2
(\bar{\sigma}_2^2-\bar{\sigma}_3^2)}  ,\qquad \phi_4^2= u_0
\frac{(1-u_1)(1-u_2)(1-u_3)}{\sigma_2^2\sigma_3^2}  ,\label{change4}
\end{gather}
such that: $0  <  u_1 <  \bar{\sigma}_2^2  <  u_2  <
\bar{\sigma}_3^2  <  u_3 <  1$, the ${\mathbb S}^3$-sphere of radius $R$ is characterized by the equation: $u_0=R^2$. The change of coordinates (\ref{change4}) corresponds to a sixteen-to-one map from the points of ${\mathbb S}^3$ out of the coordinate 3-hyperplanes $\phi_a=0$, $a=1,\dots,4$,  to the interior of the parallelepiped ${\mathbb P}_3$, in the $(u_1,u_2,u_3)$-space, def\/ined by the inequalities above. The meridian 2-spheres are eight-to-one mapped into the boundary of ${\mathbb P}_3$, where some inequalities become identities: $u_1=0$ for ${\mathbb S}^2_{I}$, $\{ u_1=\bar{\sigma}_2^2 \}   \bigcup  \{ u_2=\bar{\sigma}_2^2\}$, for ${\mathbb
S}_{II}$, and $\{ u_2=\bar{\sigma}_3^2\}   \bigcup  \{u_3=\bar{\sigma}_3^2\}$ in the ${\mathbb S}^2_{III}$ case. The one-dimensional coordinate meridians of these 2-spheres, in particular those that are the embedded sine-Gordon kink orbits are four-to-one mapped into the edges of ${\mathbb P}_3$, see Fig.~\ref{Fig07}. Finally, the North and South poles of ${\mathbb S}^3$ are two-to-one mapped into one vertex of ${\mathbb P}_3$: $v^{\pm}\equiv
(u_1,u_2,u_3)=(0,\bar{\sigma}_2^2,\bar{\sigma}_3^2)$.

The quadratic polynomial $V(\phi_1,\phi_2,\phi_3,\phi_4)$ becomes linear in the sphero-conical coordinates:
\begin{gather*}
V(u_1,u_2,u_3)=  \frac{\nu^2 R^2}{2} \left( u_1+u_2+u_3
-\bar{\sigma}_2^2-\bar{\sigma}_3^2\right),
\end{gather*}
but we write it in ``separable'' form:
\begin{gather*}
V(u_1,u_2,u_3)=\frac{\nu^2 R^2}{2}  \left( \frac{u_1
(u_1-\bar{\sigma}_2^2)(u_1-\bar{\sigma}_3^2)}{(u_1-u_2)(u_1-u_3)}\right.\\
\left.\phantom{V(u_1,u_2,u_3)=}{} +\frac{u_2
(u_2-\bar{\sigma}_2^2)(u_2-\bar{\sigma}_3^2)}{(u_2-u_1)(u_2-u_3)}+\frac{u_3
(u_3-\bar{\sigma}_2^2)(u_3-\bar{\sigma}_3^2)}{(u_3-u_1)(u_3-u_2)}\right).
\end{gather*}
The reason is that in this form it is obvious that the PDE (\ref{HJred}) reduces to the system of three uncoupled ODE that can be immediately integrated to f\/ind (\ref{ww}):
\begin{gather*}
W(u_1,u_2,u_3)  =  R^2 \left( (-1)^{\delta_1}
\sqrt{1-u_1}+(-1)^{\delta_2}  \sqrt{1-u_2}\right.\\
\left. \phantom{W(u_1,u_2,u_3)  =}{} +(-1)^{\delta_3}
\sqrt{1-u_3}\right)   , \qquad \delta_1,\delta_2,\delta_3=  0, 1   .
\end{gather*}
The Bogomolnyi splitting of the energy (\ref{bogo}) prompts the f\/irst-order equations (\ref{foe}):
\begin{gather}
\frac{dx}{U'(u_a)}=\frac{(-1)^{\delta_a}   du_a}{2u_a
(u_a-\bar{\sigma}_2^2)(u_a-\bar{\sigma}_3^2)  \sqrt{1-u_a}}
,\qquad a=1,2,3  .\label{eedd}
\end{gather}
It is convenient to pass to the variables:
\[
s_1=(-1)^{\delta_1}  \sqrt{1-u_1}  ,\qquad s_2=(-1)^{\delta_2}
\sqrt{1-u_2}  ,\qquad s_3=(-1)^{\delta_3}  \sqrt{1-u_3}
\]
because the quadratures in (\ref{eedd}) become:
\begin{gather}
\sum_{a=1}^3     \frac{ds_a}{1-s_a^2}  =  -dx  ,\qquad
\sum_{a=1}^3     \frac{ds_a}{\sigma_2^2-s_a^2}  =  -dx  ,\qquad
\sum_{a=1}^3     \frac{ds_a}{\sigma_3^2-s_a^2}  =
-dx  ,\label{orbitsm2}
\end{gather}
where only enter rational functions. The integration of (\ref{orbitsm2}) is straightforward:
\begin{gather}
 \arctanh s_1  +  \arctanh
s_2 + \arctanh s_3 =  -(x+\gamma_1), \nonumber\\
\arccoth \frac{s_1}{\sigma_2} +  \arctanh \frac{s_2}{\sigma_2} + \arctanh
\frac{s_3}{\sigma_2} =  -\sigma_2  (x+\gamma_2), \label{solb}\\
\arccoth
\frac{s_1}{\sigma_3} +  \arccoth \frac{s_2}{\sigma_3} +  \arctanh
\frac{s_3}{\sigma_3} =  -\sigma_3  (x+\gamma_3).\nonumber
\end{gather}
The addition formulas for the inverse hyperbolic functions,
and the def\/inition of ``Vieta va\-riab\-les'':
\[
A= s_1+s_2+s_3  ,\qquad  B= s_1s_2+s_1s_3+s_2s_3  ,\qquad  C=
s_1s_2s_3
\]
reduce the system of transcendent equations (\ref{solb}) to the linear system:
\begin{gather*}
 A -  t_1 B + C = t_1, \\
 \sigma_2^2 t_2
A -  \sigma_2 B + t_2 C = \sigma_2^3, \\
\sigma_3^2 A - \sigma_3 t_3 B + C = \sigma_3^3t_3,
\end{gather*}
where: $t_1=\tanh(-(x+\gamma_1))$,
$t_2=\tanh(-\sigma_2(x+\gamma_2))$,
$t_3=\tanh(-\sigma_3(x+\gamma_3))$. From the solutions $A$, $B$ and $C$ of this linear system we obtain the $s_1$, $s_2$ and $s_3$ variables as the roots of the cubic equation:
\begin{gather*}
s^3  -  A  s^2  +  B  s  -  C
= 0.
\end{gather*}
The standard Cardano parameters
\[
q=\frac{B}{3}-\frac{A^2}{9}  ,\qquad   r=\frac{1}{6} (3C-AB)+\frac{A^3}{27}   ,\qquad   \theta=\arccos
\frac{-r}{\sqrt{-q^3}}
\]
provide the three roots $s_1$, $s_2$ and $s_3$ by Cardano's formulas, and recalling that: $u_a=1-s_a^2$,
$a=1,2,3$ we f\/inally obtain:
\begin{gather}
u_1(x) =  1-\left( \frac{A}{3}   +   2 \sqrt{-q}   \cos
\frac{\theta}{3}\right)^2,\nonumber \\
u_2(x) = 1-\left( \frac{A}{3}  +
\sqrt{-q}   \left( - \cos\frac{\theta}{3}  -  \sqrt{3}   \sin \frac{\theta}{3} \right)\right)^2, \label{cardano}\\
u_3(x) = 1-\left( \frac{A}{3}  +   \sqrt{-q}   \left( - \cos
\frac{\theta}{3}  +  \sqrt{3}   \sin \frac{\theta}{3}
\right)\right)^2  ,\nonumber
\end{gather}
which in turn can be mapped back to cartesian coordinates. The explicit analytical expressions for the generic kinks are:
\begin{gather}
\begin{split}
& \phi_1(x) = (-1)^{\epsilon_1}  R  \bar{\sigma}_2\bar{\sigma}_3
\frac{ \left( \sigma_2-\sigma_3 t_2
t_3\right)\sech (x+\gamma_1)}{  \sigma_2\bar{\sigma}_3^2 -(\sigma_2^2-\sigma_3^2)
t_1t_2-\sigma_3\bar{\sigma}_2^2 t_2t_3},
\\
& \phi_2(x) = (-1)^{\epsilon_2}  R
\bar{\sigma}_2\sqrt{\sigma_2^2-\sigma_3^2}  \frac{   \left(
t_1-\sigma_3t_3\right)\sech
\sigma_2(x+\gamma_2)}{\displaystyle \sigma_2\bar{\sigma}_3^2
-(\sigma_2^2-\sigma_3^2) t_1t_2-\sigma_3\bar{\sigma}_2^2 t_2t_3},
\\
& \phi_3(x) = (-1)^{\epsilon_3}  R
\bar{\sigma}_3\sqrt{\sigma_2^2-\sigma_3^2}  \frac{ \left(
\sigma_2- t_1 t_2\right)\sech \sigma_3(x+\gamma_3) }{ \sigma_2\bar{\sigma}_3^2
-(\sigma_2^2-\sigma_3^2)
t_1t_2-\sigma_3\bar{\sigma}_2^2 t_2t_3},
\\
& \phi_4(x) = (-1)^{\epsilon_4}  R    \frac{ -\bar{\sigma}_2^2
\sigma_3  t_1-\left( \sigma_2^2-\sigma_3^2-\sigma_2\bar{\sigma}_3^2
t_1 t_2\right)   t_3}{  \sigma_2\bar{\sigma}_3^2
-(\sigma_2^2-\sigma_3^2) t_1t_2-\sigma_3\bar{\sigma}_2^2 t_2t_3},
\end{split} \label{sss}
\end{gather}
where $\epsilon_a=0,1$, $\forall\, a=1,\dots,4$. Equations (\ref{sss}) def\/ine a three-parameter family of kinks in the ${\mathbb S}^3$-sphere.

\begin{figure}[t]
\centerline{\includegraphics[height=3.5cm]{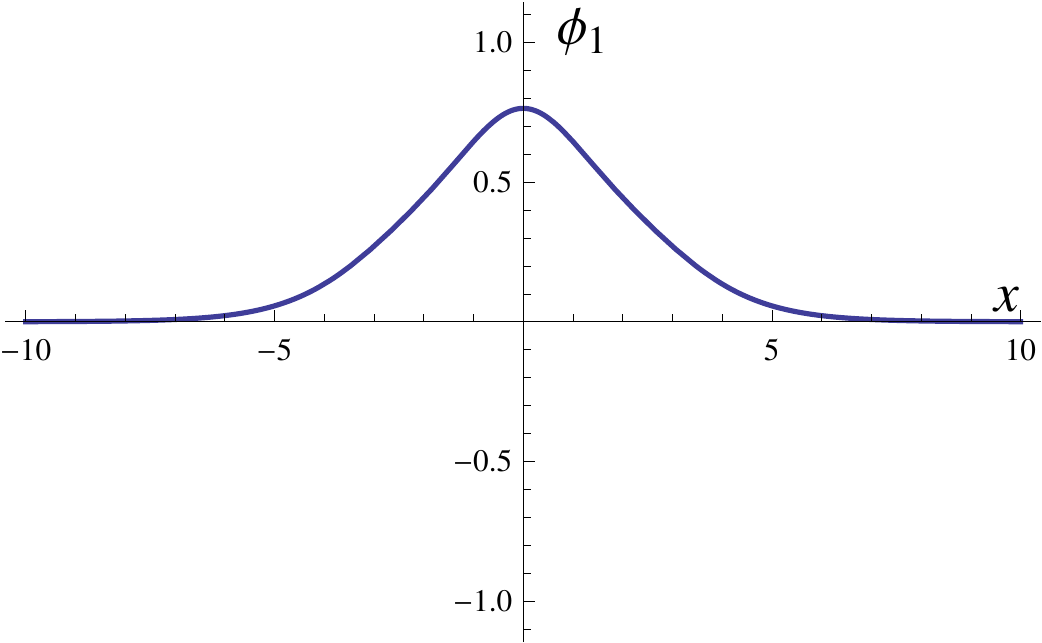}
\includegraphics[height=3.5cm]{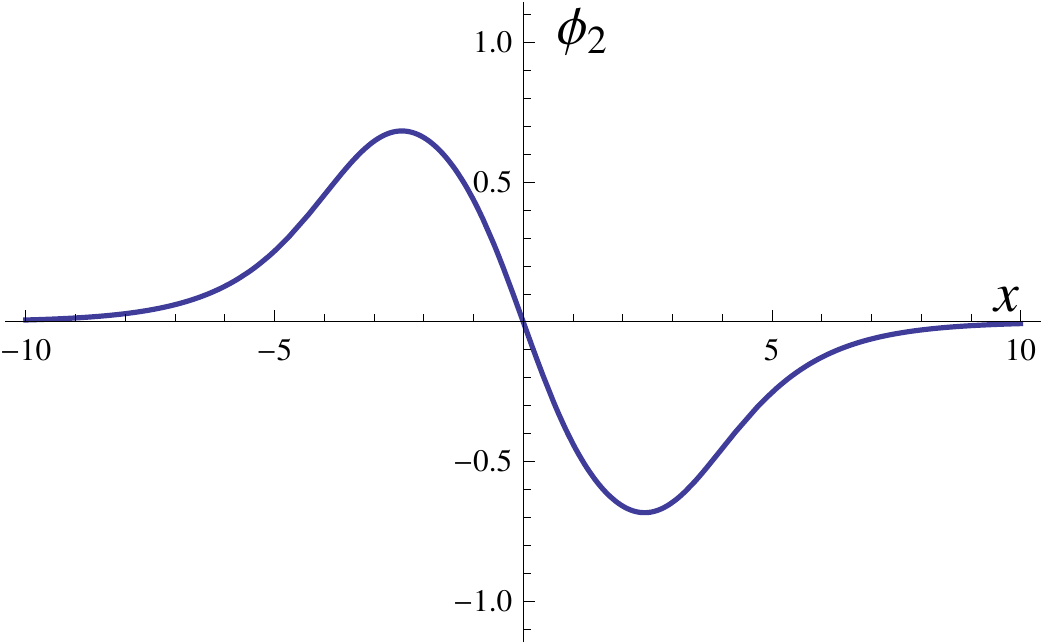}}
\smallskip
\centerline{\includegraphics[height=3.5cm]{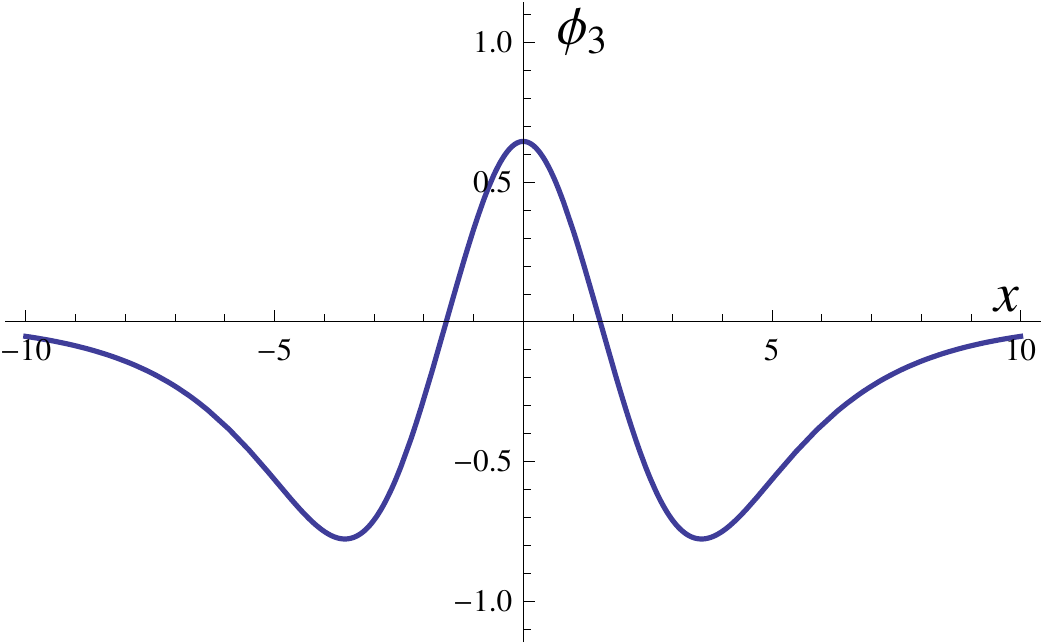}
\includegraphics[height=3.5cm]{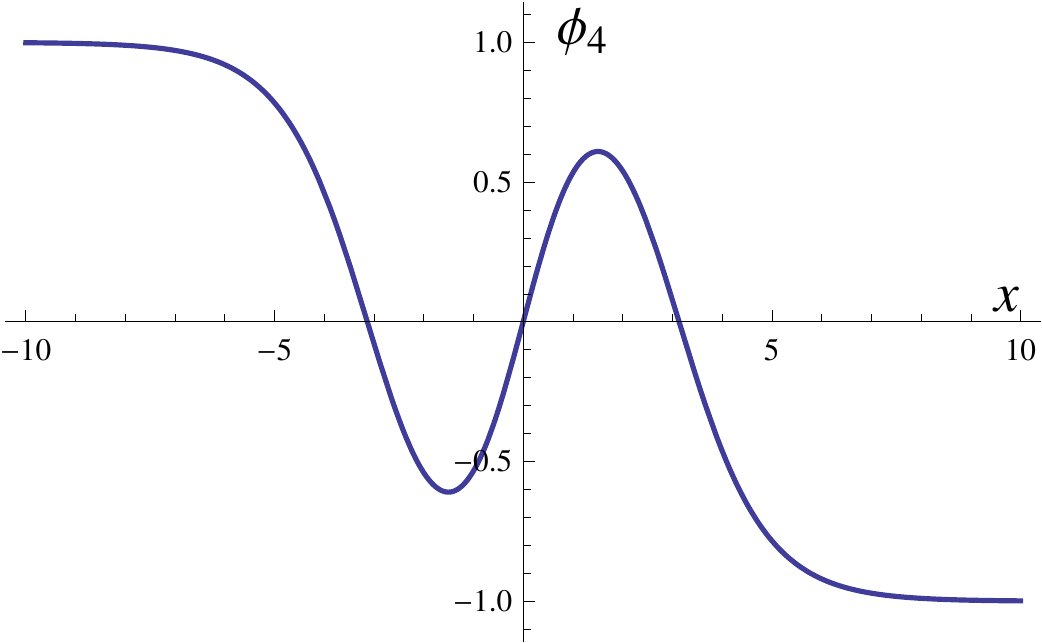}}
\caption{Graphics of the kink solutions (\ref{sss}) for the following values of the parameters: $R=1$, $\sigma_2=\frac{3}{4}$,
$\sigma_3=\frac{1}{2}$, $\gamma_1=\gamma_2=\gamma_3=0$, $\epsilon_a=0$,
$\forall\, a=1,\dots,4$. This maximally symmetric kink.}\label{Fig08}
\end{figure}

\begin{figure}[t]
\centerline{\includegraphics[height=3.5cm]{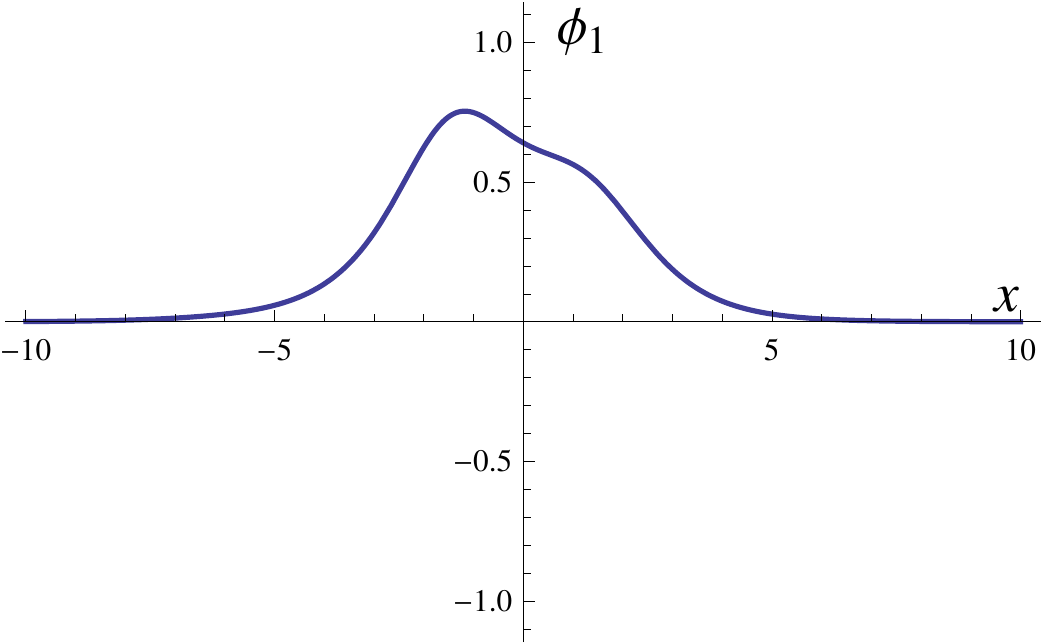}
\includegraphics[height=3.5cm]{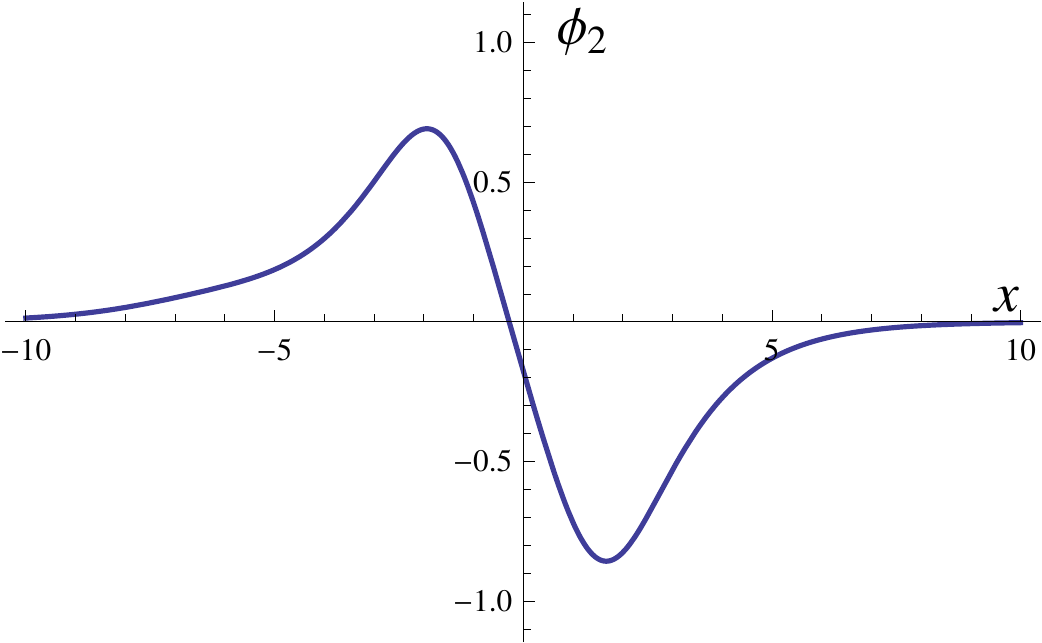}}
\smallskip
\centerline{\includegraphics[height=3.5cm]{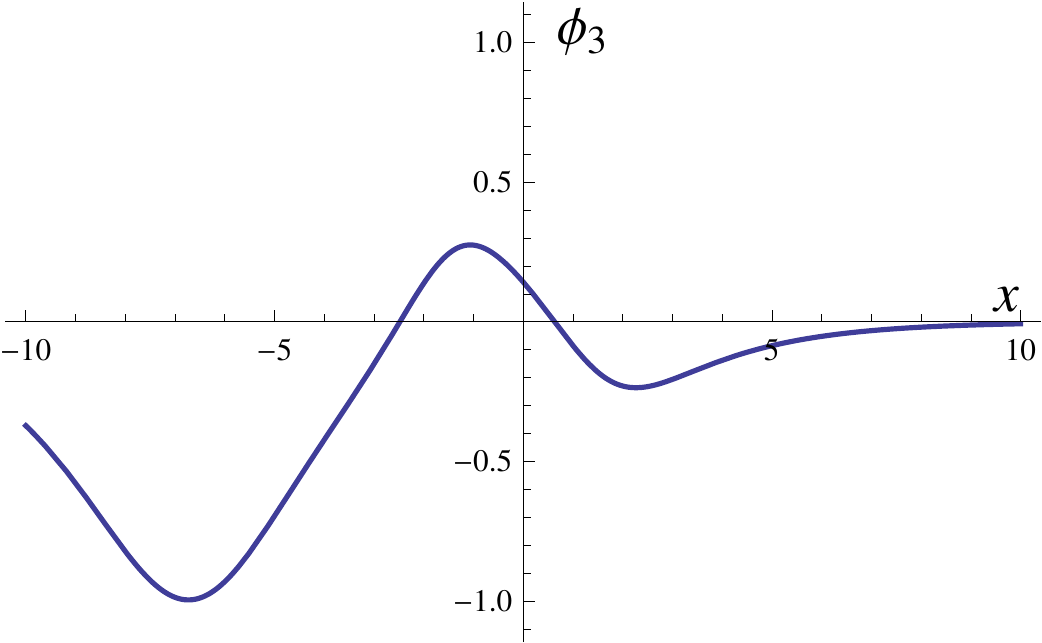}
\includegraphics[height=3.5cm]{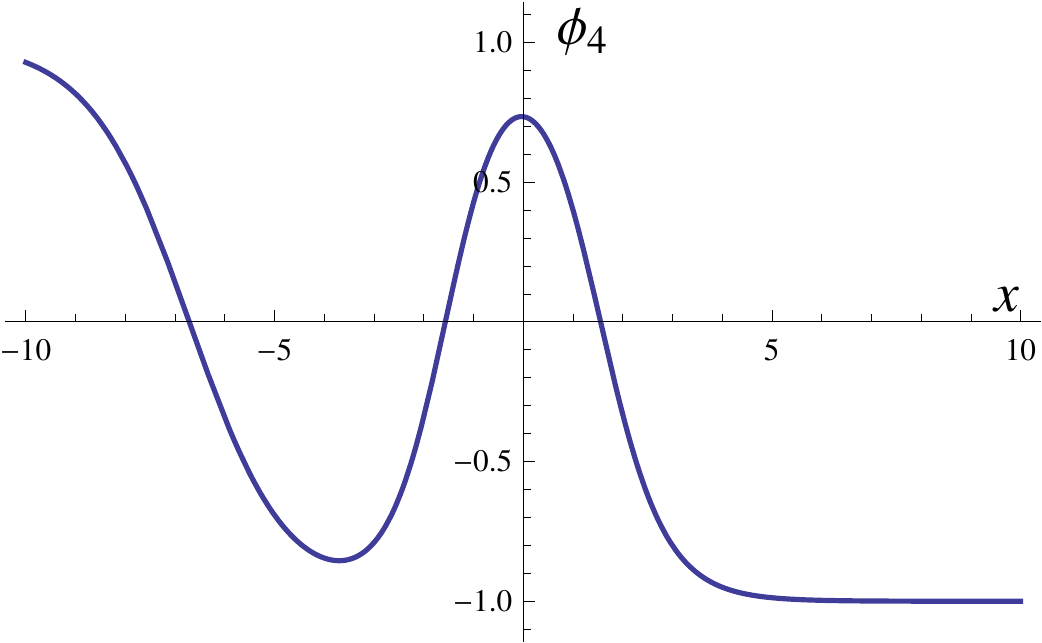}}
\caption{Graphics of (\ref{sss}) for: $R=1$, $\sigma_2=\frac{3}{4}$,
$\sigma_3=\frac{1}{2}$, $\gamma_1=0.8$, $\gamma_2=1$, $\gamma_3=4$,
$\epsilon_a=0$, $\forall\, a=1,\dots,4$.}\label{Fig09}
\end{figure}

\subsection{The structure of the moduli space of kinks}

In the Figs.~\ref{Fig08} and \ref{Fig09} two generic kink prof\/iles have been plotted for representative values of the parameters.

A glance to the plots of $\phi_4$ as function of $x$ shows that the kinks are topological interpolating between the North and the South poles. In fact, all the generic kinks are topological. The ana\-lysis of the asymptotic behavior of the fourth f\/ield component in (\ref{sss}) determines the topological character of all the solitary waves solutions. The generic kinks belong to either the ${\cal C}_{\rm NS}$ or the ${\cal C}_{\rm SN}$ sectors of the conf\/iguration space. The limits
\begin{gather*}
\begin{split}
& \lim_{x\to -\infty}\left(
\phi_1(x),\phi_2(x),\phi_3(x),\phi_4(x)\right)  =  (0,0,0,
(-1)^{\epsilon_4}  R),\\
&\lim_{x\to \infty}\left(
\phi_1(x),\phi_2(x),\phi_3(x),\phi_4(x)\right)  =  (0,0,0,
-(-1)^{\epsilon_4}  R)
\end{split}
\end{gather*}
show that $\epsilon_4=0$ implies that the $TK$ kink family lies in ${\cal C}_{\rm NS}$, whereas $\epsilon_4=1$ sends the $TK$ family to live in ${\cal C}_{\rm SN}$. The other sign options, $(-1)^{\epsilon_1}$, $(-1)^{\epsilon_2}$ and $(-1)^{\epsilon_3}$, determine the three-dimensional hemispheres accommodating the dif\/ferent $TK$ kink orbits, as well as the kink/antikink character of the solutions, in a smooth generalization of the $N=2$ $NTK$ kinks.

\begin{figure}[t]
\centerline{\includegraphics[height=3.5cm]{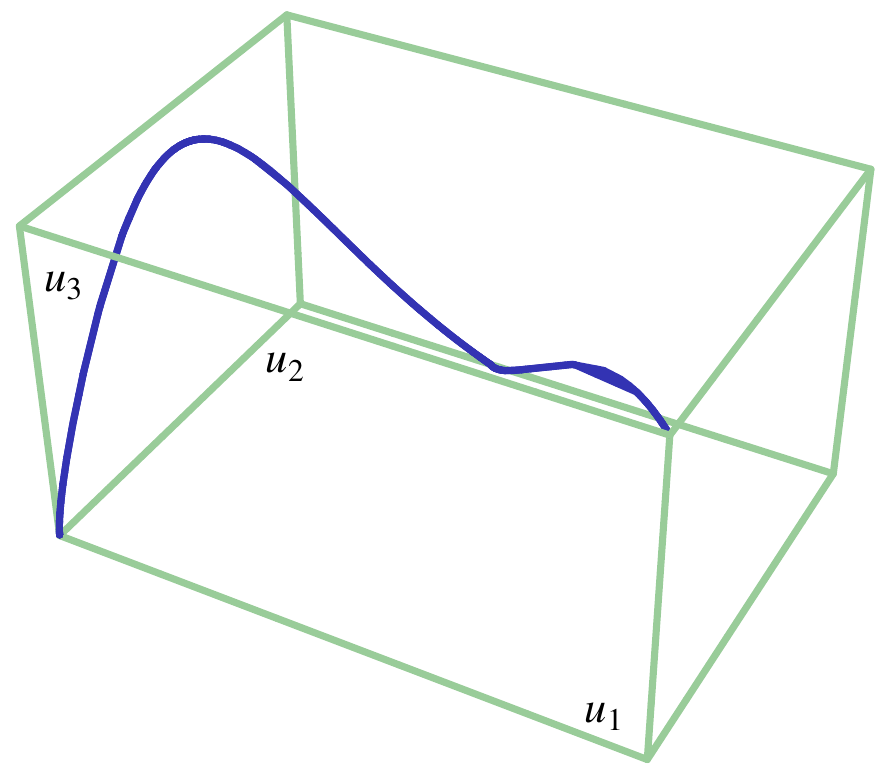}\quad
\includegraphics[height=3.5cm]{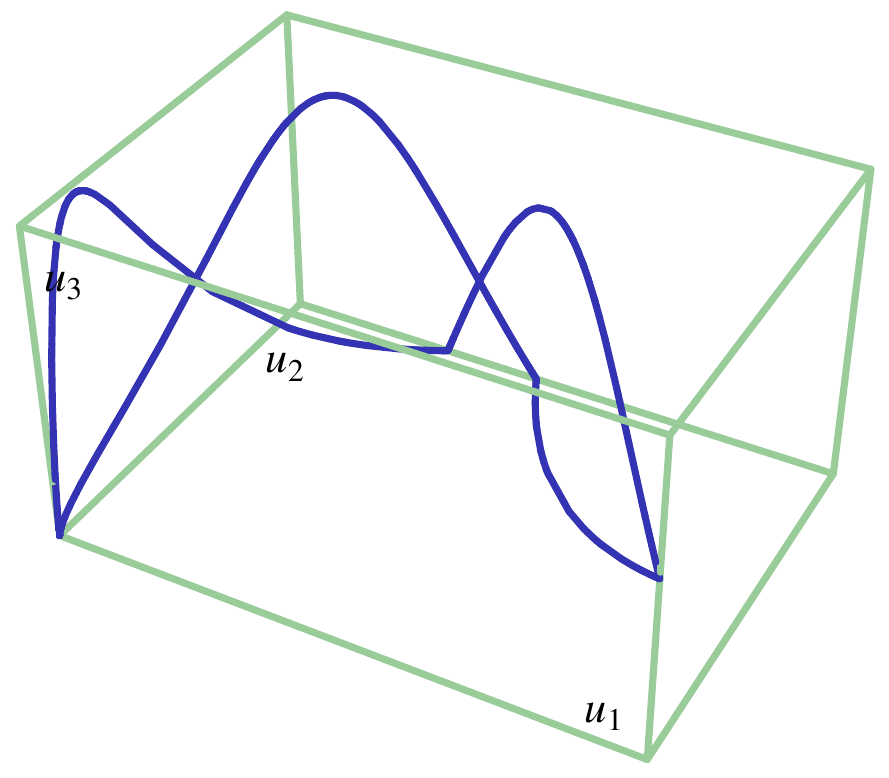}\quad \includegraphics[height=3.5cm]{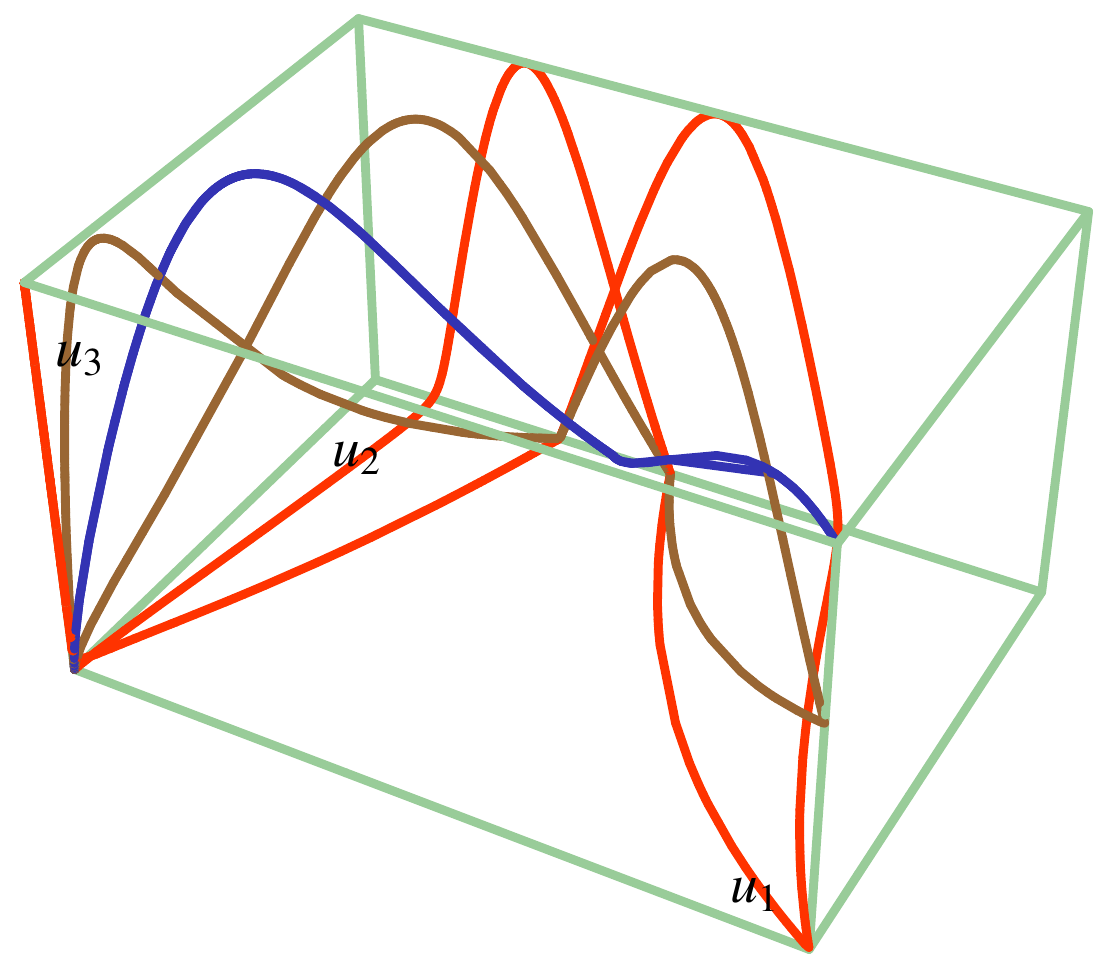}}
\caption{$TK$ orbit corresponding to: $\gamma_1=\gamma_2=\gamma_3=0$ (left).  The case: $\gamma_1=0.8$, $\gamma_2=2$ and $\gamma_3=2$ (middle). Three orbits corresponding to dif\/ferent values of the parameters (right).}
\label{Fig10}
\end{figure}

\begin{figure}[t]
\centerline{\includegraphics[height=3cm]{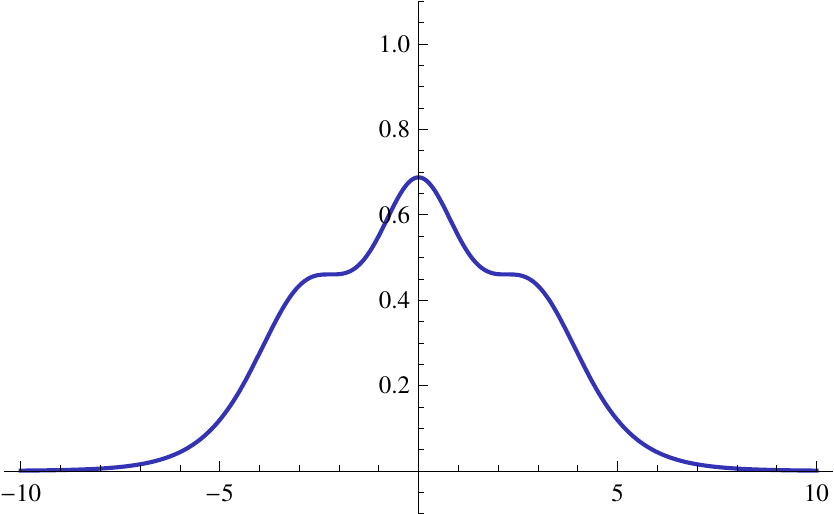}
\includegraphics[height=3cm]{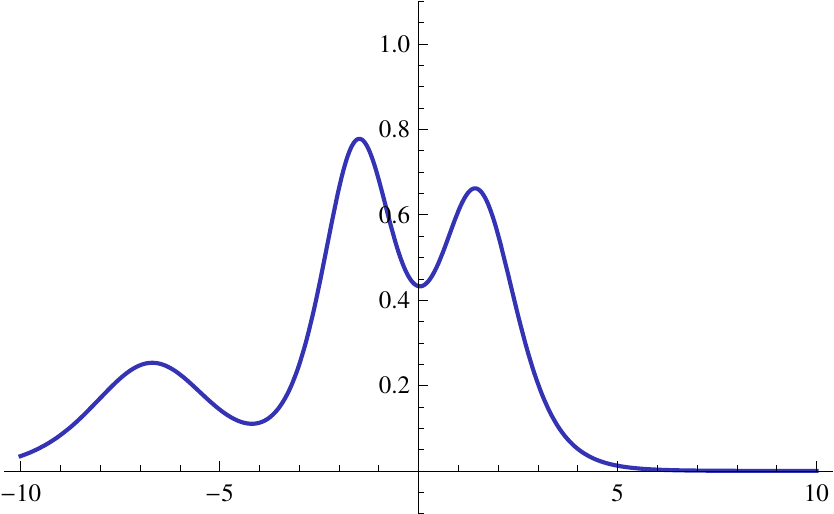}  \includegraphics[height=3cm]{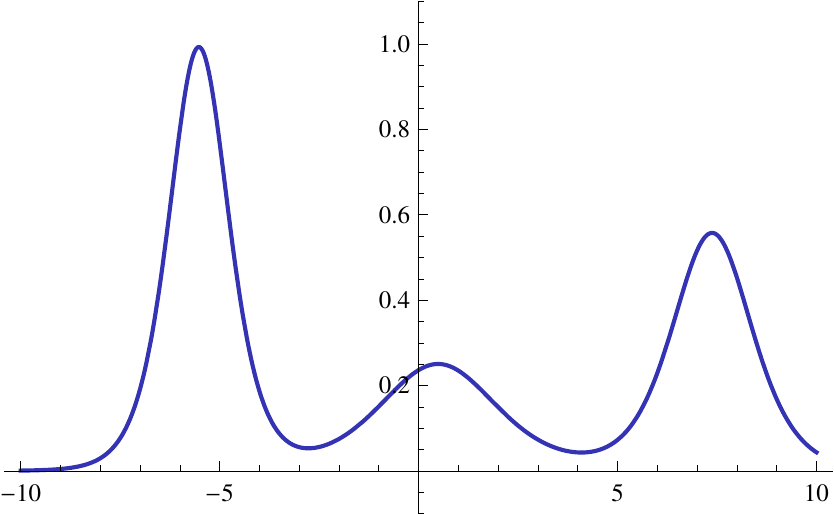}}
\caption{Kink energy densities for dif\/ferent values of the constants
$\gamma_1$, $\gamma_2$ and $\gamma_3$.}\label{Fig11}
\end{figure}

It is possible to draw the kink orbits in the ${\mathbb P}_3$
parallelepiped using the solutions (\ref{cardano}) in
sphero-conical coordinates. In the Fig.~\ref{Fig10} (right) several $TK$ orbits are drawn together. The energy of all the $TK$ kinks is:
\[
E_{TK} = 2\nu R^2(1+\sigma_2+\sigma_3)  =  E_{K_1}+E_{K_2}+E_{K_3}  .
\]
There is a new kink energy sum rule specif\/ically arising in the $N=3$ model. In the Fig.~\ref{Fig11} it can be viewed how the energy density ref\/lects this energy sum
rule: the energy density of a given member of the $TK$ family is
a ``composition'' of the energy densities of the sine-Gordon kinks $K_1$,
$K_2$ and~$K_3$, see Fig.~\ref{Fig11} (middle) and (right). This structure is completely similar to the kink manifold of the f\/ield theoretical model with target space f\/lat ${\mathbb R}^3$ and the Garnier system as mechanical analogue problem, see \cite{Nonlinearity1} and \cite{Nonlinearity}. In fact, the equations are identical, even thought in that case the problem is separable using elliptical coordinates in ${\mathbb R}^3$, \cite{Nonlinearity1, Nonlinearity}.
The calculation of the limits of the $TK$ family when the parameters
$\gamma_1$, $\gamma_2$ and $\gamma_3$ tend to $\pm \infty$ is not a
simple task, but it is possible to show that all the possibilities
arising from the energy sum rule are obtained for the dif\/ferent
limits (see \cite{Nonlinearity} for details; the model is dif\/ferent but the equations are the same and thus the
analytical calculation of the limiting cases coincides). In the Fig.~\ref{Fig10} (right), for instance, a $TK$ kink orbit (in red color) close to the limiting combination~$K_3$ plus
a~member of the $NTK_{III}$ family, is plotted.

Finally, the structure of the kink manifolds for the $N=2$ and $N=3$ models
allows us to foresee that in higher $N$, the generic kinks will be topological if $N$ is an odd number, and
non-topological for even $N$. In fact, the kink energy sum rule will
establish that the energy of a generic solitary wave is the
sum of the energies of the $N$ sine-Gordon kinks embedded in the model.

\section{Kink stability}\label{section7}

The analysis of the kink stability requires the study of the small
f\/luctuations around kinks. For simplicity, we will only present the
explicit results for the topological kinks of the ${\mathbb S}^2$
model, and also a geometrical analysis in terms of Jacobi f\/ields
that allows us to demonstrate the instability of the non-topological
kinks. The generalization to the arbitrary $N$ case is
straightforward.

\subsection[Small fluctuations on topological kinks, $K_1$ and $K_2$]{Small f\/luctuations on topological kinks, $\boldsymbol{K_1}$ and $\boldsymbol{K_2}$}

Taking into account the explicit expressions of the $K_1$ and $K_2$
sine-Gordon kinks of the ${\mathbb S}^2$ model, we will use spherical
coordinates in ${\mathbb S}^2$ in order to alleviate the notation.
Thus we introduce the coordinates:
\[
\phi=R \sin \theta^1 \cos \theta^2  ,\qquad   \phi_2=R\sin \theta^1\sin
\theta^2  ,\qquad   \phi_3=R\cos \theta^1
\]
with  $\theta^1\in[0,\pi]$, $\theta^2\in[0,2\pi)$. The metric tensor
over ${\mathbb S}^2$ is written now as: $ds^2=R^2
d\theta^1d\theta^1+R^2 \sin^2\theta^1d\theta^2d\theta^2$. The
associated Christof\/fel symbols and Riemann curvature tensor
components will be:
\begin{gather*}
\Gamma^1_{22}=-\frac{1}{2}\sin 2\theta^1, \qquad
\Gamma^2_{12}=\Gamma^2_{21}=\cotan \theta^1 ,\\
R^1_{212}=
-R^1_{122}={\rm sin}^2\theta^1,\qquad   R^2_{121}= -R^2_{211}=1.
\end{gather*}

The analysis of small f\/luctuations around a kink $\theta_K(x)\equiv
(\theta_K^1(x),\theta_K^2(x))$ is determined by the second-order
Hessian operator:
\begin{gather}
\Delta_K\eta=-\big(
\nabla_{\theta_K^\prime}\nabla_{\theta_K^\prime}\eta+
R(\theta^\prime_K,\eta)\theta^\prime_K +\nabla_\eta {\rm
grad}\, V\big), \label{eq:hess}
\end{gather}
i.e.\ the geodesic deviation operator plus the Hessian of the
potential energy density, valuated at the kink. $\eta$ denotes the
perturbation around the kink. Let $\theta(x)$ denote the deformed
trajectory, $\theta(x)=\theta_K(x)+\eta(x)$, with
$\eta(x)=(\eta^1(x),\eta^2(x))$, then we introduce the following
contra-variant vector f\/ields along the kink trajectory, $\eta ,
\theta^\prime_K \in \Gamma(T{\mathbb S}^2\left|_K\right.)$:
\[
\eta(x)=\eta^1(x)\frac{\partial}{\partial\theta^1}+\eta^2(x)\frac{\partial}{\partial\theta^2} ,\qquad
\theta^\prime_K(x)=\theta_K^{\prime  1}
\frac{\partial}{\partial\theta^1}+\theta_K^{\prime
2}\frac{\partial}{\partial\theta^2}.
\]
We will use standard notation for covariant derivatives and Riemann
tensor:
\[ \nabla_{\theta_K^\prime} \eta=\big(\eta^{\prime
i}(x)+\Gamma^i_{jk}\eta^j\theta_K^{\prime
k}\big)\frac{\partial}{\partial\theta^i} ,\qquad
R(\theta^\prime_K,\eta)\theta^\prime_K=\theta_K^{\prime
i}\eta^j(x)\theta_K^{\prime
k}R^l_{ijk}\frac{\partial}{\partial\theta^l}.
\]
The geodesic deviation operator and the Hessian of the potential
read:
\begin{gather*}
\frac{D^2\eta}{dx^2}+
R(\theta^\prime_K,\eta)\theta^\prime_K=\nabla_{\theta_K^\prime}\nabla_{\theta_K^\prime}\eta+
R(\theta^\prime_K,\eta)\theta^\prime_K  ,  \\
  \nabla_\eta \, {\rm
grad}\, V=\eta^i\left(\frac{\partial^2
V}{\partial\theta^i\partial\theta^j}-\Gamma^k_{ij}\frac{\partial
V}{\partial\theta^k}\right)g^{jl}\frac{\partial}{\partial\theta^l}
\end{gather*}
evaluated at $\theta_K(x)$. In sum, second-order kink f\/luctuations
are determined by the operator:
\begin{gather}
\Delta_K\eta  =  -\left(
\frac{D^2\eta}{dx^2}+R(\theta^\prime_K,\eta)\theta^\prime_K
+\nabla_\eta \, {\rm grad}\, V\right)    \nonumber\\
\phantom{\Delta_K\eta }{}  =
-\left(\frac{d^2\eta^1}{dx^2}-\cos 2\theta_K^{1}  \left[
\big(\theta_K^{\prime
2}\big)^2+\sigma^2+\bar{\sigma}^2\cos^2\theta_K^{2}\right] \eta^1 -\sin
2\theta_K^{1}  \theta_K^{\prime  2}  \frac{d\eta^2}{dx}\right.
\nonumber \\   \left.
\phantom{\Delta_K\eta = }{}
- \left[ \big(1+\cos
2\theta_K^{1}\big)\theta_K^{\prime  1}\theta_K^{\prime  2}+\frac{\sin
2\theta_K^{1}}{2}\left(\theta_K^{\prime\prime
2}-\bar{\sigma}^2\frac{\sin
2\theta_K^{2}}{2}\right)\right] \eta^2\right)\frac{\partial}{\partial\theta^1}\nonumber \\
\phantom{\Delta_K\eta = }{}
-\left(2 \cotan \theta_K^{1}  \theta_K^{\prime  2}
\frac{d\eta^1}{dx}+\big(\cotan \theta_K^{1}  \theta_K^{\prime\prime
2}-\theta_K^{\prime  1}\theta_K^{\prime
2}\big)\eta^1+\frac{d^2\eta^2}{dx^2} \right. \nonumber \\
\left.
\phantom{\Delta_K\eta = }{}
 +  2
\cotan \theta_K^{1}  \theta_K^{\prime  1}  \frac{d\eta^2}{dx
}+\big(\cotan\theta_K^{1}    \theta_K^{\prime\prime
1}-(\theta_K^{\prime  1})^2-\cos^2\theta_K^{1}
\big(\theta_K^{\prime  2}\big)^2\big)\right) \frac{\partial}{\partial\theta^2}.
\label{eq:hess2}
\end{gather}

 {\bf The spectrum of small f\/luctuations around $\boldsymbol{K_2/K_2^*}$
kinks.}  The $K_2/K_2^*$ kink solu\-tions~(\ref{k1k2sG}), (\ref{k1k2sG2})
are written, in spherical coordinates, as follows:
\begin{gather*}
  \theta_{K_2}^1   =  2 \arctan e^{\pm \sigma (x-x_0)}   ,\qquad
\theta_{K_2}^2   =  \frac{\pi}{2}, \\   \theta_{K_2^*}^1   =  2
\arctan e^{\pm \sigma (x-x_0)}   ,\qquad \theta_{K_2^*}^2   =
\frac{3\pi}{2},
\end{gather*}
where the $(\pm)$ sign determine the kink/antikink choice, and we
will take $x_0=0$ for simplicity. Plugging these $K_2/K_2^*$ kink
solutions in~(\ref{eq:hess2}), we obtain the dif\/ferential operator
acting on the second-order f\/luctuation operator around the
$K_2/K_2^*$ kinks:
\begin{gather}
\Delta_{K_2}\eta=\left[
-\frac{d^2\eta^1}{dx^2}+\left(\sigma^2\!-\frac{2\sigma^2}{{\rm
cosh}^2\sigma
x}\right)\!\eta^1\right]\frac{\partial}{\partial\theta^1} +\left[
-\frac{d^2\eta^2}{dx^2}+2\sigma{\rm tanh}\sigma
x\frac{d\eta^2}{dx}+\bar{\sigma}^2\eta^2\right]\frac{\partial}{\partial\theta^2}.\!\!\!
 \label{eq:hess1}
\end{gather}
We know from classical dif\/ferential geometry that this expression
can be simplif\/ied if one uses a parallel basis along the kink
trajectory. Therefore let us consider the equations of parallel
transport along $K_2$ kinks: $\nabla_{\theta_{K_2}^\prime} v=0$,
where $v$ is the vector-f\/ield:
$v(x)=v^1(x)\frac{\partial}{\partial\theta^1}+v^2(x)\frac{\partial}{\partial\theta^2}$.
These equations are written explicitly as:
\[
\frac{dv^i}{dx}+\Gamma^i_{jk}\bar{\theta}^{\prime j} v^k=0
\Rightarrow     \left\{\begin{array}{ll} \displaystyle \frac{dv^1}{dx}=0 &
  \Rightarrow    \    v^1(x)=1, \vspace{2mm}\\
  \displaystyle  \frac{dv^2}{dx}+\sigma\frac{\cotan
   (2 \arctan  e^{\sigma x})}{\cosh \sigma x}   v^2=0  & \Rightarrow \
    v^2(x)=\cosh \sigma x. \end{array}\right.
\]
Therefore the vector-f\/ields: $v_1=\frac{\partial}{\partial
\theta^1}$ and $v_2(x)= \cosh \sigma x   \frac{\partial}{\partial
\theta^2}$ form a frame $\left\{ v_1,v_2\right\}$ in
$\Gamma(T{\mathbb S}^2|_{K_2})$ parallel to the $K_2$ kink in which
(\ref{eq:hess1}) reads as follows:
\begin{gather*}
\Delta_{K_2}\eta=\Delta_{K_2^*}\eta=\left[
-\frac{d^2\bar{\eta}^1}{dx^2}+\left(\sigma^2-\frac{2\sigma^2}{\cosh^2\sigma
x}\right)\bar{\eta}^1\right]  v_1+ \left[
-\frac{d^2\bar{\eta}^2}{dx^2}+\left(1-\frac{2\sigma^2}{\cosh^2\sigma
x}\right)\bar{\eta}^2\right]  v_2,
\end{gather*}
where $\eta=\bar{\eta}^1   v_1+\bar{\eta}^2   v_2$,
$\eta^1=\bar{\eta}^1$, and $\eta^2=\cosh \sigma x  \bar{\eta}^2$.
Thus the second-order f\/luctuation operator, expressed in the
parallel frame, is a diagonal matrix of transparent P\"osch--Teller
Schr\"odinger operators with very well known spectra. In fact, in
the $v_1=\frac{\partial}{\partial\theta^1}$ direction we f\/ind the
Schr\"odinger operator governing sine-Gordon kink f\/luctuations, as could be
expected. The novelty is that we f\/ind another P\"osch--Teller
potential of the same type in the
$v_2=\frac{\partial}{\partial\theta^2}$ direction, orthogonal to the
kink trajectory. The spectra is given by:
\begin{itemize}\itemsep=0pt
\item $v_1$ direction.  There is a bound state of zero eigenvalue and
a continuous family of positive eigenfunctions:
\begin{gather*}
\bar{\eta}^1_0(x)=\sech \sigma x  ,\qquad \varepsilon^{(1)}_0=0, \\
 \bar{\eta}^1_k(x)=e^{i k \sigma x}({\rm tanh}\sigma x - i k)
  ,\qquad   \varepsilon^{(1)}(k)=\sigma^2\left(k^2+1\right).
\end{gather*}

\item $v_2=\cosh\sigma x\frac{\partial}{\partial\theta^2}$ direction.
The spectrum is similar but the bound state corresponds to a~positive eigenvalue:
\begin{gather*}
\bar{\eta}^2_{1-\sigma^2}(x)=\sech \sigma x  ,\qquad
\varepsilon^{(2)}_{1-\sigma^2}=1-\sigma^2>0,\\
\bar{\eta}^2_k(x)=e^{i k \sigma x}({\rm tanh}\sigma x - i k)   ,\qquad
\varepsilon^{(2)}(k)=\sigma^2k^2+1.
\end{gather*}
Because there are no f\/luctuations of negative eigenvalue, the
$K_2/K_2^*$ kinks are stable.
\end{itemize}

 {\bf The spectrum of small f\/luctuations around $\boldsymbol{K_1/K_1^*}$
kinks.} A similar procedure shows that the $K_1$ kink/antikink are
unstable. The $K_1/K_1^*$ kink solutions (\ref{k1k2sG}),
(\ref{k1k2sG2}), in spherical coordinates, are as follows:
\begin{gather*}
 \theta_{K_1}^1   =  2 \arctan e^{\pm (x-x_0)}   ,\qquad
\theta_{K_1}^2   =  0,\\   \theta_{K_1^*}^1   =  2 \arctan
e^{\pm (x-x_0)}   ,\qquad \theta_{K_1^*}^2   =  \pi,
\end{gather*}
where we f\/ind again that the $(\pm)$ sign determine the
kink/antikink choice, and it will be taken $x_0=0$ for simplicity.
By inserting the $K_1/K_1^*$ solutions in~(\ref{eq:hess}) the
second-order f\/luctuation operator around the $K_1/K_1^*$ kinks is
found:
\begin{gather}
\Delta_{K_1}\eta=\Delta_{K_1^*}\eta=\left[
-\frac{d^2\eta^1}{dx^2}+\left(1-\frac{2}{{\rm cosh}^2
x}\right)\eta^1\right]\frac{\partial}{\partial\theta^1}\nonumber\\
\phantom{\Delta_{K_1}\eta=\Delta_{K_1^*}\eta=}{} +\left[
-\frac{d^2\eta^2}{dx^2}+2\tanh
x\frac{d\eta^2}{dx}-\bar{\sigma}^2\eta^2\right]\frac{\partial}{\partial\theta^2}. \label{eq:hess2a}
\end{gather}
An analogous calculation lead to the parallel frame along the
$K_1/K_1^*$ kink orbits:
\[
\{w_1,w_2 \}\in\Gamma\big(T{\mathbb S}^2\big|_{K_1}\big),\qquad
w_1= \frac{\partial}{\partial \theta^1}  ,\qquad w_2(x)= \cosh x
\frac{\partial}{\partial \theta^2}.
\]
And thus (\ref{eq:hess2a}) becomes:
\begin{gather}
\Delta_{K_1}\eta=\Delta_{K_1^*}\eta=\left[
-\frac{d^2\tilde{\eta}^1}{dx^2}+\left(1-\frac{2}{\cosh^2
x}\right)\tilde{\eta}^1\right] w_1\nonumber\\
\phantom{\Delta_{K_1}\eta=\Delta_{K_1^*}\eta=}{}
+\left[
-\frac{d^2\tilde{\eta}^2}{dx^2}+\left(\sigma^2-\frac{2}{\cosh^2
x}\right)\tilde{\eta}^2\right] w_2 \label{eq:hess22}
\end{gather}
with $\eta=\tilde{\eta}^1 w_1+\tilde{\eta}^2 w_2$,
$\eta^1=\tilde{\eta}^1$, $\eta^2=\cosh x\tilde{\eta}^2$.

Again, the second-order f\/luctuation operator (\ref{eq:hess22}) is a
diagonal matrix of transparent P\"osch--Teller operators. In this
case, there is a bound state of zero eigenvalue and a continuous
family of positive eigenfunctions starting at the threshold
$\varepsilon^{(1)}(0)=1$ in the
$w_1=\frac{\partial}{\partial\theta^1}$ direction:
\[
\tilde{\eta}^1_0(x)=\frac{1}{\cosh x}  ,\qquad
\varepsilon^{(1)}_0=0, \qquad \tilde{\eta}^1_k(x)=e^{i k
x}(\tanh x - i k)  ,\qquad  \varepsilon^{(1)}(k)=\left(k^2+1\right).
\]
In the $w_2(x)=\cosh x\frac{\partial}{\partial\theta^2}$ direction,
the spectrum is similar but the eigenvalue of the bound state is
negative, whereas the threshold of this branch of the continuous
spectrum is $\varepsilon^{(2)}(0)=\sigma^2$:
\begin{gather*}
\tilde{\eta}^2_{\sigma^2-1}(x)=\frac{1}{\cosh x}  ,\qquad
\varepsilon^{(2)}_{\sigma^2-1}=\sigma^2-1<0,\\
\tilde{\eta}^2_k(x)=e^{i k x}(\tanh x - i k)  ,\qquad
\varepsilon^{(2)}(k)=k^2+\sigma^2.
\end{gather*}
Therefore, $K_1/K_1^*$ kinks are unstable and a Jacobi f\/ield for
$k=i\sigma$ arises: $\tilde{\eta}^2_J(x)=e^{\sigma x}(\tanh
x-\sigma)$, $\varepsilon^{(2)}_J=0$.

Finally, it is remarkable that this results can be reproduced
without dif\/f\/iculty in the cases $N>2$ by using hyper-spherical
coordinates in ${\mathbb S}^N$. The only stable kink that is present
in the model is the sine-Gordon $K_N$ kink, that corresponds to the minimum
of the kink energies.

\subsection[The stability of the $NTK$ kinks]{The stability of the $\boldsymbol{NTK}$ kinks}

The analysis of the spectrum of small f\/luctuations around the kinks
of the $NTK$ family is not an easy task, and thus we will proceed by
another way, see \cite{Nonlinearity, ItoT}. Having obtained
explicit expressions $\vec{\Phi}^{NTK}(x; \gamma, \bar{\gamma})$ for
the $NTK$ family of solutions~(\ref{sol2}), we know from classical
Morse--Jacobi theory about conjugate points, that the vector-f\/ields
$\frac{\partial \vec{\Phi}^{NTK}}{\partial \gamma}$ and
$\frac{\partial \vec{\Phi}^{NTK}}{\partial \bar{\gamma}}$ are Jacobi
f\/ields along the $NTK$ trajectories, i.e.\ zero-modes of the spectrum
of small f\/luctuations around the solutions. In fact, taking into
account that the $\gamma$ parameter simply determines the ``center''
of the kink, it is easy to conclude that $\frac{\partial
\vec{\Phi}}{\partial \gamma}$ is tangent to the kink trajectories,
and thus only $\frac{\partial \vec{\Phi}^{NTK}}{\partial
\bar{\gamma}}$ is a genuine (i.e. orthogonal to the kink-orbits)
Jacobi f\/ield for this family of solutions \cite{Nonlinearity}.

\begin{figure}[t]
\centerline{\includegraphics[height=3.5cm]{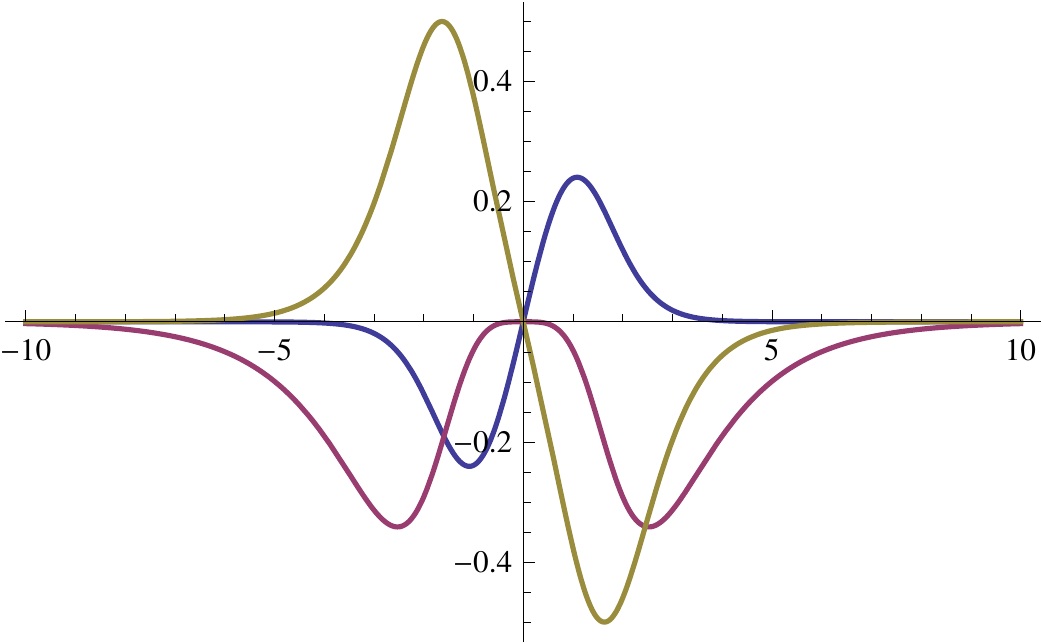}\quad
\includegraphics[height=3.5cm]{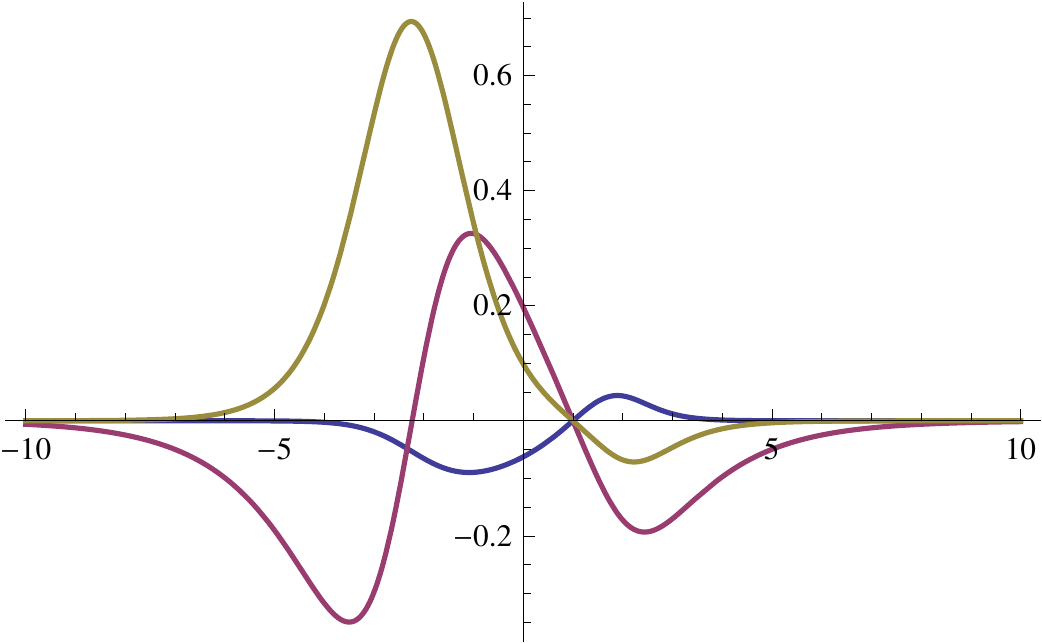}}
\caption{Components of the Jacobi f\/ields (\ref{jacobif}) for
$\sigma=0.7$, $\epsilon_1=\epsilon_2=\epsilon_3=0$, and
$\gamma=\bar{\gamma}=0$ (left), $\gamma=1$, $\bar{\gamma}=2$ (right).}\label{Fig12}
\end{figure}

By derivation on (\ref{sol2}) with respect to $\bar{\gamma}$
parameter, it is obtained the Jacobi f\/ield:
\begin{gather}
J^{NTK}(x;\gamma,\bar{\gamma})   =   \frac{\partial
\vec{\Phi}^{NTK}}{\partial \bar{\gamma}}  =
\frac{R\sigma\bar{\sigma}\sech (\sigma(x-\gamma+\bar{\gamma}))
\tanh (x-\gamma)}{(1+\sigma \tanh(x-\gamma)
\tanh (\sigma(x-\gamma+\bar{\gamma}))^2}    \nonumber \\
\phantom{J^{NTK}(x;\gamma,\bar{\gamma})   =}{}
\times \left( (-1)^{\epsilon_1}\sigma \sech(x-\gamma)
\sech(\sigma(x-\gamma+\bar{\gamma}))  \frac{\partial \  }{\partial
\phi_1}\right. \nonumber  \\
\phantom{J^{NTK}(x;\gamma,\bar{\gamma})   =}{}
 +  (-1)^{\epsilon_2} \left( \sigma
\tanh(x-\gamma)+\tanh(\sigma(x-\gamma+\bar{\gamma}))\right)
\frac{\partial \ }{\partial \phi_2}\nonumber \\   \left.
\phantom{J^{NTK}(x;\gamma,\bar{\gamma})   =}{}
-
(-1)^{\epsilon_3}\bar{\sigma} \sech(\sigma(x-\gamma+\bar{\gamma}))
\frac{\partial \ }{\partial \phi_3}\right).\label{jacobif}
\end{gather}
By a direct calculation, it can be checked that
\[
\lim_{x\to \pm \infty} J^{NTK}(x;\gamma,\bar{\gamma})   =
0,\qquad J^{NTK}(\gamma;\gamma,\bar{\gamma})  =  0
\]
and thus the point $x=\gamma$ in the kink trajectory, i.e.\
$\vec{\Phi}^{NTK}(\gamma; \gamma,\bar{\gamma})  =
((-1)^{\epsilon_1}R \sigma,0, (-1)^{\epsilon_3}R \bar{\sigma})$ is a
conjugate point of the corresponding minima S (or N depending on
the choice of $\epsilon_3$). In Fig.~\ref{Fig12} there are depicted the
components of (\ref{jacobif}) for several dif\/ferent cases.

The existence of a conjugate point establishes, according with Morse
theory, that these kink solutions are not stable under small
perturbations. In fact, the Morse index theorem states that the
number of negative eigenvalues of the second order f\/luctuation
operator around a given orbit is equal to the number of conjugate
points crossed by the orbit \cite{ItoT}. The reason is that the
spectrum of the Schr\"odinger operator has in this case an
eigenfunction with as many nodes as the Morse index, the Jacobi
f\/ield, whereas the ground state has no nodes. The Jacobi f\/ields of
the $NTK$ orbits cross one conjugate point, their Morse index is
one, and the $NTK$ kinks are unstable.

Finally, it is possible to extend these results to the $N=3$ case,
where two Jacobi f\/ields appear, and the instability of the generic
$TK$ family is showed. The procedure is the same, but the extension
of the expressions is considerably bigger, thus we will not include
this calculation here.

\subsection*{Acknowledgements}

We are very grateful to J. Mateos Guilarte for informative and
illuminating conversations on several issues concerning this work.
We also thank the Spanish Ministerio de Educaci\'on y Ciencia and
Junta de Castilla y Le\'on for partial support under grants
FIS2006-09417 and GR224.

\pdfbookmark[1]{References}{ref}
\LastPageEnding

\end{document}